\documentclass{aastex631}

\usepackage{amsmath,amstext}
\usepackage{tikz}
\usepackage{booktabs}
\usepackage{graphicx}
\usepackage{epstopdf}
\usepackage{soul}

\newcommand{\nfrag}{$N_{frag}$}

\newcommand{\paperi}{Paper I}
\newcommand{\paperii}{Paper II}

\shortauthors{Humphries, Morgan, Kuridze}

\begin{document}

\title{An in-depth analysis of quiet-Sun IRIS Brightenings}
%\footnote{Released on June, 10th, 2019}

\correspondingauthor{Humphries, Morgan, Kuridze}
\email{ldh6@aber.ac.uk, hum2@aber.ac.uk, dak21@aber.ac.uk}

\author[0000-0002-0786-7307]{Ll\^yr Dafydd Humphries}
\affiliation{Aberystwyth University \\
Faculty of Business and Physical Sciences\\
Aberystwyth, Ceredigion, SY23 3FL, Wales, UK}

\author[0000-0002-0786-7307]{Huw Morgan}
\affiliation{Aberystwyth University \\
Faculty of Business and Physical Sciences\\
Aberystwyth, Ceredigion, SY23 3FL, Wales, UK}

\author[0000-0003-2760-2311]{David Kuridze}
\affiliation{National Solar Observatory, 3665 Discovery Drive, Boulder, CO 80303, USA}
\affiliation{Aberystwyth University \\
Faculty of Business and Physical Sciences\\
Aberystwyth, Ceredigion, SY23 3FL, Wales, UK}
\affiliation{Georgian National Astrophysical Observatory, Abastumani, 0301, Georgia}

\begin{abstract}

Small-scale brightenigs are ubiquitous, dynamic and energetic phenomena found in the chromopshere. 
An advanced filter-detection algorithm applied to high-resolution observations from the Interface Region Imaging Spectrograph enables the detection of these brightenings close to the noise level. This algorithm also tracks the movement of these brightenings and extracts their characteristics. This work outlines the results of an in-depth analysis of a quiet-Sun dataset including a comparison of a brighter domain - associated with a super-granular boundary - to the quiescent inter-network domains. Several characteristics of brightenings from both domains are extracted and analysed, providing a range of sizes, durations, brightness values, travel distances, and speeds. The ``Active" quiet-Sun events tend to travel shorter distances and at slower speeds along the plane-of-sky than their ``True" quiet-Sun counterparts. These results are consistent with the magnetic field model of super-granular photospheric structures and the magnetic canopy model of the chromosphere above. Spectroscopic analyses reveal that BPs demonstrate blue-shift (as well as some bi-directionality) and that they may rise from the chromosphere into the TR. We believe these bright points to be magnetic in nature, are likely the result of magnetic reconnection, and follow current sheets between magnetic field gradients, rather than along magnetic field lines themselves. 
    
\end{abstract}
 
\section{Introduction}

Small-scale, bright, energetic events are ubiquitous across the solar surface, regardless of solar cycle or regional activity. These brightenings, or small flare-like events, can be seen in imaging and spectroscopic data, and occur at a large range of energies. The chromosphere is typically host to these phenomena, the magnetic and thermal structure of which is complex
\citep{Aschwanden_2008, Klimchuk_2014}. Some of these events's origins are theorised to begin within or below the photosphere, such as spicules, which are generally considered to be the result of photospheric acoustic waves that cascade into a series of rebounding shock fronts in the chromsphere above \citep{Rae_1982, Rouppe_van_der_Voort_2003, Stein_1972}. Similar small-scale phenomena have been observed from several instruments, including (but not limited to) the Interface Region Imaging Spectrograph \citep[IRIS,][]{pontieu}, the Atmospheric Imaging Assembly aboard the Solar Dynamic Observatory \citep[AIA, SDO,][]{pesnell2012}, the Reuven Ramaty High Energy Solar Spectroscopic Imager \citep[RHESSI,][]{Lin_2002} and Hinode \citep{Kosugi_2007}. 

Another example of a phenomenon that may be intimately linked with the photosphere and observed by IRIS are ``campfires" - typically classified as small-scale coronal brightenings that are found 1000-5000km above the photosphere \citep{Zhukov_2021} and have been observed in several instances, such as 
%rooted in regions of weak magnetic field locations \citep{Carvajal_in_proceedings_2022}, 
at the edge of a chromospheric network clump or lane \citep{Berghmans_2021}, above photospheric magnetic neutral lines \citep{Panesar_2021}, and rooted at chromospheric network boundaries \citep{Berghmans_2021, Zhukov_2021}. 
They can take many forms, including jet-/loop-like and dot-like structures, as well as more complex configurations. Other small-scale energetic events of interest are UV bursts that exhibit brightness intensity enhancements of an order of magnitude and strong plasma flows \citep{pontieu}, IRIS bombs and Ellerman bombs, which typically demonstrate asymmetrical spectral results and chromospheric/photospheric brightenings \citep{Georgoulis_2002,Peter_2014}.

Microflares are defined as small-scale flares, characteristically with energies of order $10^{-6}$ times that of large, typical flares \citep{Hannah_2011}. They are believed to be the result of the same magnetic reconnection mechanism that give rise to their large-scale counterparts, whereby magnetic fields in the solar surface are displaced by photospheric plasma convection, resulting in a braiding of field lines and the formation of small-scale current layers, and the eventual explosive release of accumulated magnetic energy \citep{Bogachev_2020}. These scaled-down eruptive events, like large flares, are also characterised by temperatures that can reach that of the corona \citep[$\sim10$ MK][]{Reale_2019, Glesener_2020}, accompanying electron accelerations \citep{Cooper_2021, Wright_2017}, and their potential contribution to coronal heating \citep{Hudson_1991, Hannah_2008}. 
Even fainter, smaller analogues, typically speculated to lie below the resolution of most instruments and dubbed ``nanoflares", have also been suggested as a mechanism by which energy is transferred to the corona. This is postulated to occur via rapid and ubiquitous magnetic energy dissipation as a result of random, continuous motions of magnetic footpoints, and are considered as the most basic impulsive energy unit of large scale flares and/or as a constant, ever-present mass of small-scale eruptions \citep{Parker_1988, Cargill_1996, Galsgaard_1996, Priest_et_al_2002}. Several simulation-related analyses imply that this is indeed possible, such as \cite{Testa_2020}'s findings; modelling of nanoflare-heated loops reproduce characteristics of short-lived ($\sim25$s) Si IV brightenings - observed by IRIS - that have coronal heating counterparts observed in the AIA 94 \AA\ channel, suggesting that particle acceleration can occur even in very small-scale reconnection events.

It is clear, then, that the chromosphere and photosphere exhibit a variety of small-scale, energetic, ubiquitous, elusive, and insufficiently understood phenomena. Their relationship with the corona, if any, is unclear, and observations of such phenomena remain difficult-, particularly regarding their automatic detection when both the scale and luminosity of such events approach the noise level. 

This study expands on the work conducted by \cite{Humphries_2021_a} (hereafter \paperi) and \cite{Humphries_2021_b} (hereafter \paperii) by applying the brightening detection/extraction method to another quiet-Sun (QS) dataset. Their detection code has been tuned to detect events that are as faint and small as possible, and the viability of their method has been tested on synthetic data (see \paperi\ for details).

\begin{figure*}[t]
    \centering
    (a)\includegraphics[width=0.28\textwidth]{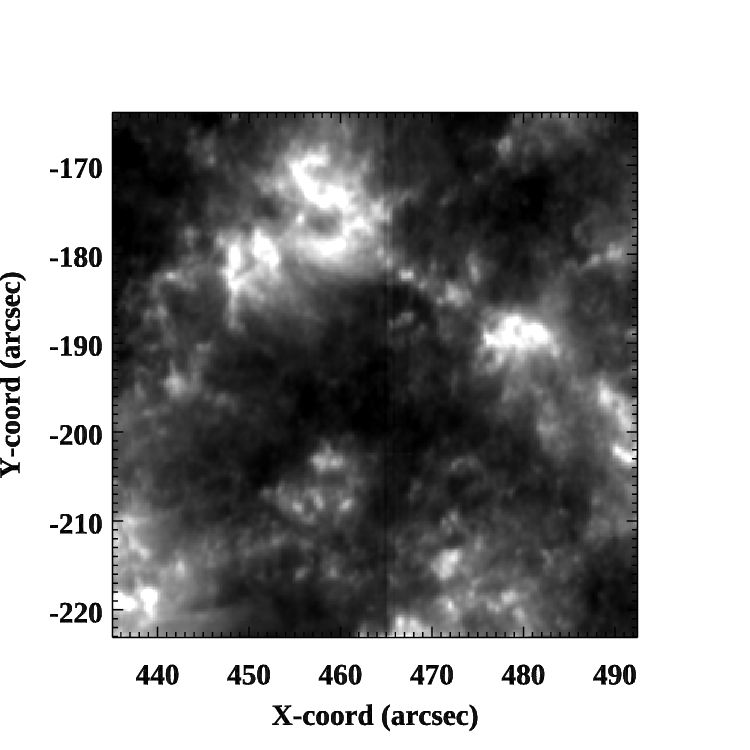}
    (b)\includegraphics[width=0.28\textwidth]{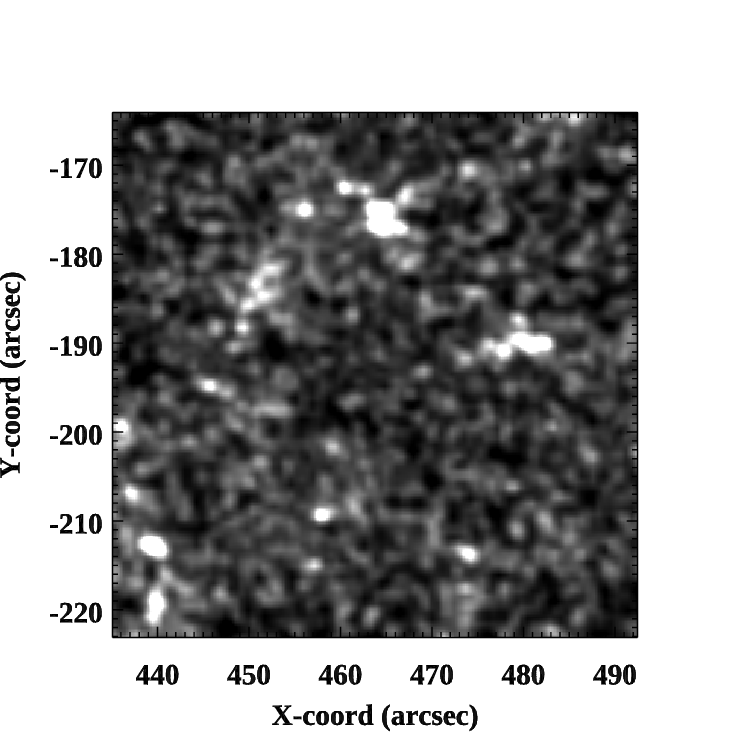}
    (c)\includegraphics[width=0.28\textwidth]{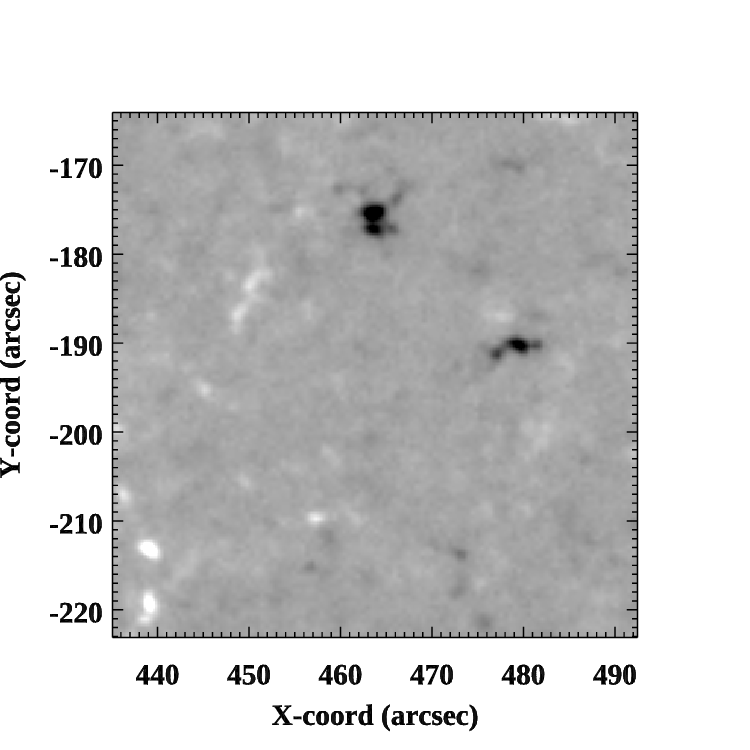}
    \includegraphics[trim={10cm 0 0 0},clip, width=0.057\textwidth]{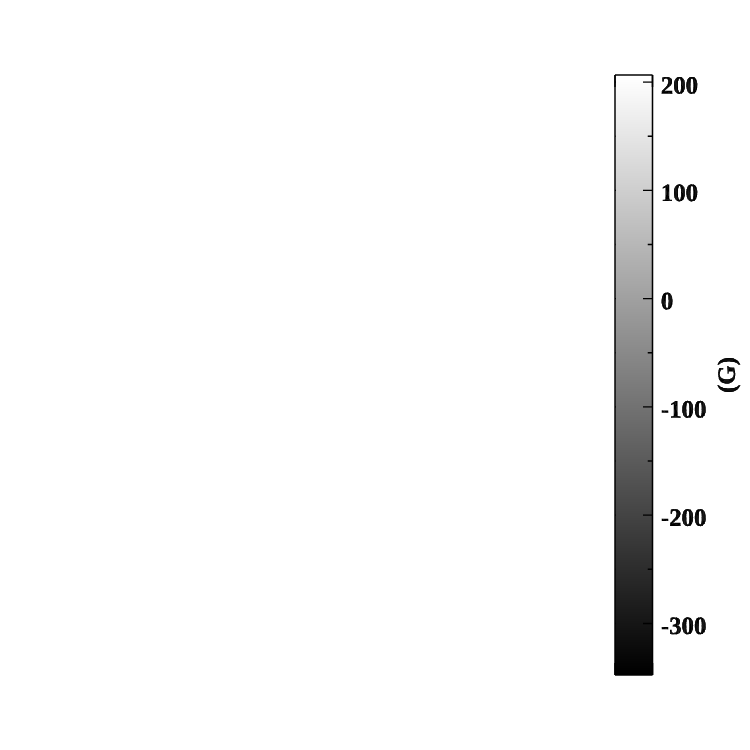}
    \caption{Time-averaged image (across $\sim3.2$ hours) of the (a) 1400 \AA\ 2013-09-26 dataset, (b) context 1700 \AA\ image from AIA, and (c) context HMI magnetogram, all of which are co-aligned for the same FOV. Colourbar corresponds to HMI image (measured in Gauss).}
    \label{fig:FOV_comp}
\end{figure*}

The purpose of this study is to present a statistical analysis of the behaviour of small-scale bright points (BPs) within different domains of an IRIS QS region as observed in a single wavelength channel. BPs detected in relatively ``active" domains of the QS field-of-view (FOV) are compared with their counterparts detected in relatively quiescent domains, providing insight into their nature. The abundance of information returned on over 12,600 detections can provide some understanding of the quiet-Sun nature and ubiquity of these events, as well as the premise for a detailed comparison with other solar regions in the future.

Section \ref{sec:obs_and_meth} provides a brief summary of the observations and regions of interest (ROIs), as well as an overview of the detection and analysis methodology. Results are presented and discussed in sections \ref{sec:results} and \ref{sec:disc}, and conclusions are provided in section \ref{sec:con}.

\begin{figure*}[t]
    \centering
    \includegraphics[width=0.6\textwidth]{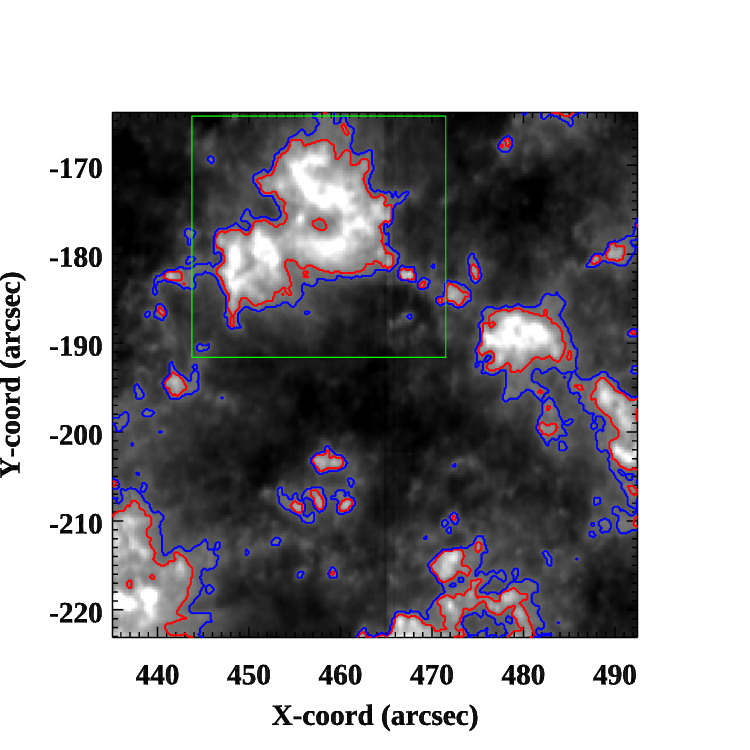}
    \caption{IRIS 1400 \AA\ image with domain contours. Domains that lie within red contours are considered ``Active" QS. Domains outside the blue contours are considered ``True" QS. Domains that lie between contour colours are ignored. The green box indicates a more focused FOV - see text for details.}
    \label{fig:FOV_w_contours_and_box}
\end{figure*}

\section{Observations \& Method} \label{sec:obs_and_meth}

\subsection{Observations}\label{sec:sub_obs}

We select an IRIS 1400 \AA\ slit-jaw image (SJI) dataset, consisting of a series of sit-and-stare slit-jaw images, a 0.17$\arcsec \times 0.17\arcsec$ pixel spatial scale, and an approximately consistent temporal cadence of $\sim27$s. Figure \ref{fig:FOV_comp}a shows a time-averaged image (over $\sim3.2$ hours) of the region of interest (ROI) with the presence of some brighter features. Alongside are (b) AIA 1700 \AA\ and (c) Helioseismic Magnetic Imager \citep[HMI,][]{pesnell2012} images of the same FOV from SDO. The original dataset - available on the Lockheed Martin Solar and Astrophysics Laboratory (LMSAL) archive - is centred at $X = 458\arcsec$, $Y = -194\arcsec$, with a $65\arcsec\times69\arcsec$ FOV, while cropped images - as those seen throughout this paper - are centred at $X = 468\arcsec$, $Y = -194\arcsec$, with a $57\arcsec\times59\arcsec$ FOV, having removed empty margins typical of IRIS datasets. The observation begins Sept. 26th 2013 at 21:26 UT and ends on Sept. 27th 2013 at 00:37 UT, while the original dataset provided by LMSAL begins at 20:09. However, high-energy particle bombardment from the South Atlantic Anomaly (SAA) necessitates discarding the first $\sim170$ frames.

We focus solely on the Si {\sc{iv}} 1400 \AA\ channel for several reasons: this is the most common dataset available for sit-and-stare IRIS data, it allows a direct comparison with previous studies, it facilitates future comparative studies, and was deemed appropriate due to the complexity of this study.

The IRIS dataset is processed using standard IRIS level 2 pipeline procedures, such as dark current subtraction, geometric/orbital calibration, and flat fields. Any over-saturated pixels are ignored by the filtering/detection method.
Contrary to \paperi\ and \paperii, the central slit and its surrounding pixels remain intact and are not treated as missing data: it was found that these darker pixels do not affect the filtering process by any significant measure. Sub-pixel image translations due to solar rotation have been corrected using a Fourier Local Correlation Tracking method \citep{fw}. 

Spectroscopic data is also available from several wavelength channels, including (but not limited to) C {\sc{ii}} (1334 \AA\ and 1336 \AA), Si {\sc{iv}} (1394 \AA\ and 1403 \AA), Cl {\sc{i}} 1352 \AA\, and O {\sc{i}} 1356 \AA. These are extracted using default IDL functions as well the \textit{IRIS\_GETWINDATA} and \textit{IRIS\_AUTO\_FIT} routines. \textit{IRIS\_AUTO\_FIT} fits the spectrosocpic data according to a single (or double) Gaussian where appropriate for each image pixel. The standard central peak positions for each wavelength are used as reference and calibration wavelengths.

\begin{splitdeluxetable*}{lcccccBccccBccccccc}
\tablewidth{\textwidth}
\tablecaption{Comparison of BP properties detected within the Full FOV, AQS domain, and TQS domain of the IRIS 2013-09-26 20:09:36 dataset. All BPs detected are $N_{frag}\le2$. 
\label{tab:data_AQS_vs_TQS}}
\tablehead{
\colhead{Dataset} &
\colhead{BPs} & 
\colhead{\% of} & 
\colhead{Event} & 
\colhead{Average area} & 
\colhead{Area OES} &
\colhead{Speed OES} & 
\colhead{Overall Speed} & 
\colhead{Total Brightness} & 
\colhead{Maximum Brightness} & 
\colhead{Duration} & 
\colhead{Overall} & 
\colhead{AOM} &
\colhead{Overall Distance} &
\colhead{Distance OES} \\
\colhead{} & 
\colhead{} & 
\colhead{Detections} & 
\colhead{Density}& 
\colhead{min/$\tilde{x}$/$\bar{x}$/max} & 
\colhead{min/$\tilde{x}$/$\bar{x}$/max} & 
\colhead{min/$\tilde{x}$/$\bar{x}$/max} & 
\colhead{min/$\tilde{x}$/$\bar{x}$/max} & 
\colhead{min/$\tilde{x}$/$\bar{x}$/max} & 
\colhead{min/$\tilde{x}$/$\bar{x}$/max} & 
\colhead{min/$\tilde{x}$/$\bar{x}$/max} & 
\colhead{AOM} & 
\colhead{OES} & 
\colhead{min/$\tilde{x}$/$\bar{x}$/max} & 
\colhead{min/$\tilde{x}$/$\bar{x}$/max} \\
\colhead{} & 
\colhead{} & 
\colhead{(\%)} & 
\colhead{($\rm arcsec^{-2}~s^{-1}$)} &
\colhead{($\rm arcsec^{2}$)} & 
\colhead{($\rm arcsec^{2}$)} &
\colhead{($\rm km~s^{-1}$)} & 
\colhead{($\rm km~s^{-1}$)} & 
\colhead{($\rm DN~s^{-1}$)} & 
\colhead{($\rm DN~s^{-1}$)} & 
\colhead{($\rm s$)} & 
\colhead{($\rm ^{\circ}$)} &  
\colhead{($\rm ^{\circ}$)} &
\colhead{(arcsec)} &
\colhead{(arcsec)} &
}
\startdata
Full FOV & 12246 & 100 & $3.142\times10^{-4}$ & 0.174 / 0.565 / 0.608 / 2.213 & 0.028 / 0.553 / 0.617 / 2.878 & 0.004 / 5.032 / 6.138 / 64.789  & 0.184 / 5.083 / 5.107 / 16.428 & 20 / 1002 / 1484 / 22977 & 3 / 21 / 23 / 211 & 80 / 134 / 172 / 991 & -1.770 / -2.612 & -1.091 / -2.676 & 0.006 / 0.683 / 0.766 / 4.976 & 0.000 / 0.189 / 0.231 / 2.431\\
AQS & 1380 & 11.269	& $2.428\times10^{-4}$ & 0.191 / 0.543 / 0.509 / 2.034 & 0.028 / 0.546 / 0.498 / 2.269 & 0.004 / 4.896 / 3.873 / 60.474 & 0.184 / 4.079 / 3.728 / 14.819	& 70 / 2286 / 1466 / 22977 & 3 / 30 / 27 / 159 & 80 / 196 / 161 / 830 & 4.631 / 3.520 & 0.669 / 2.394 & 0.009 / 0.637 / 0.571 / 2.526 & 0.000 / 0.184 / 0.145 / 2.269 \\
TQS & 9484 & 77.446 & $2.979\times10^{-4}$ & 0.174 / 0.619 / 0.576 / 2.214 & 0.028 / 0.631 / 0.581 / 2.878 & 0.032 / 6.390 / 5.291 / 54.178 & 0.380 / 5.297 / 5.265 / 16.429 & 20 / 1335 / 934 / 13210 & 3 / 22 / 20 / 211 & 80 / 168 / 134 / 991 & -3.162 / -1.486 & -3.105 / -1.174 & 0.001 / 0.791 / 0.701 / 4.976 & 0.001 / 0.240 / 0.199 / 2.033\\
\enddata
\tablecomments{AQS - ``Active" QS region. TQS - ``True" QS region. See text for details. $\tilde{x}$ denotes median. $\bar{x}$ denotes mean.}
\end{splitdeluxetable*}

\begin{figure*}[t]
    \centering
    (a)\includegraphics[width=0.3\textwidth]{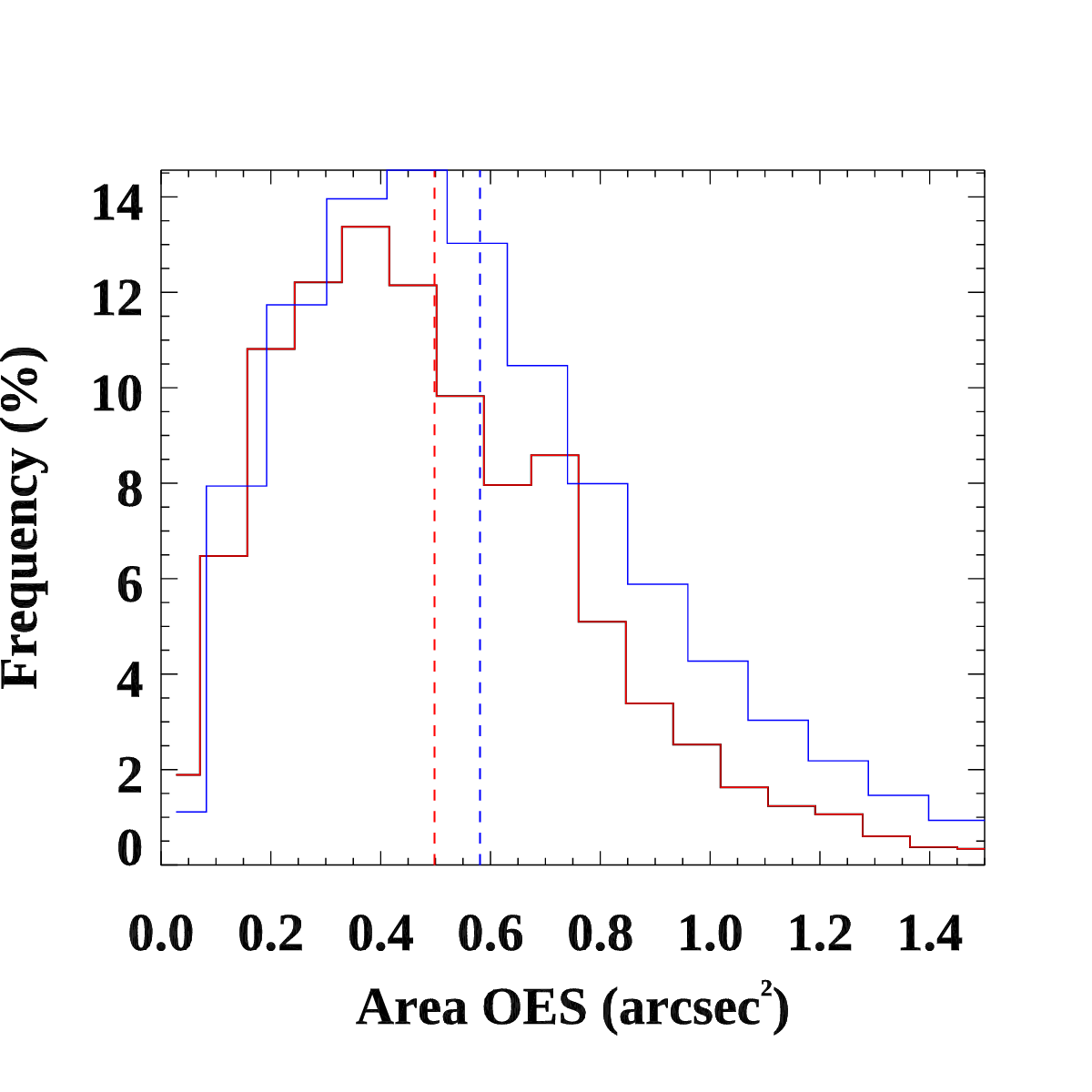}
    (b)\includegraphics[width=0.3\textwidth]{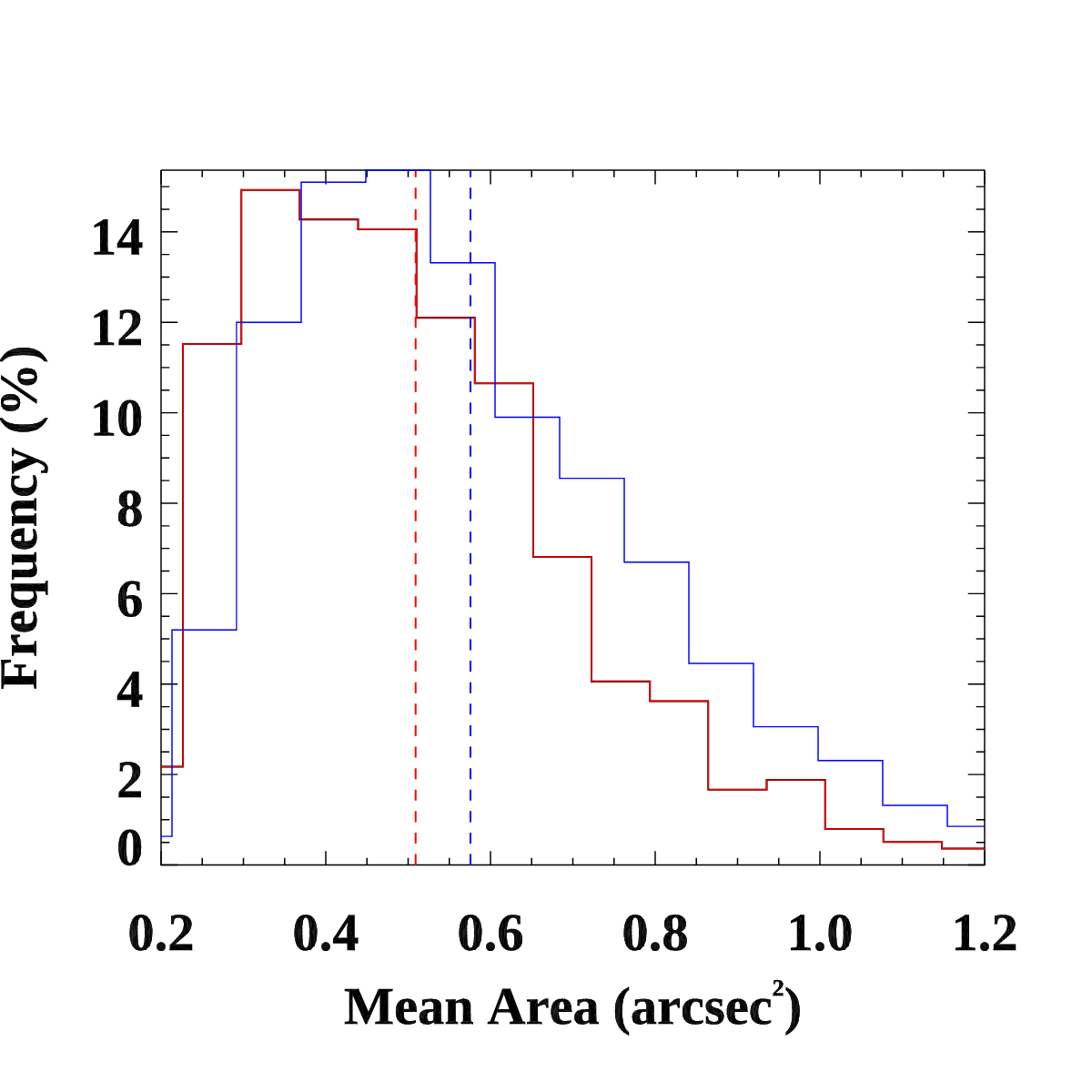}
    (c)\includegraphics[width=0.3\textwidth]{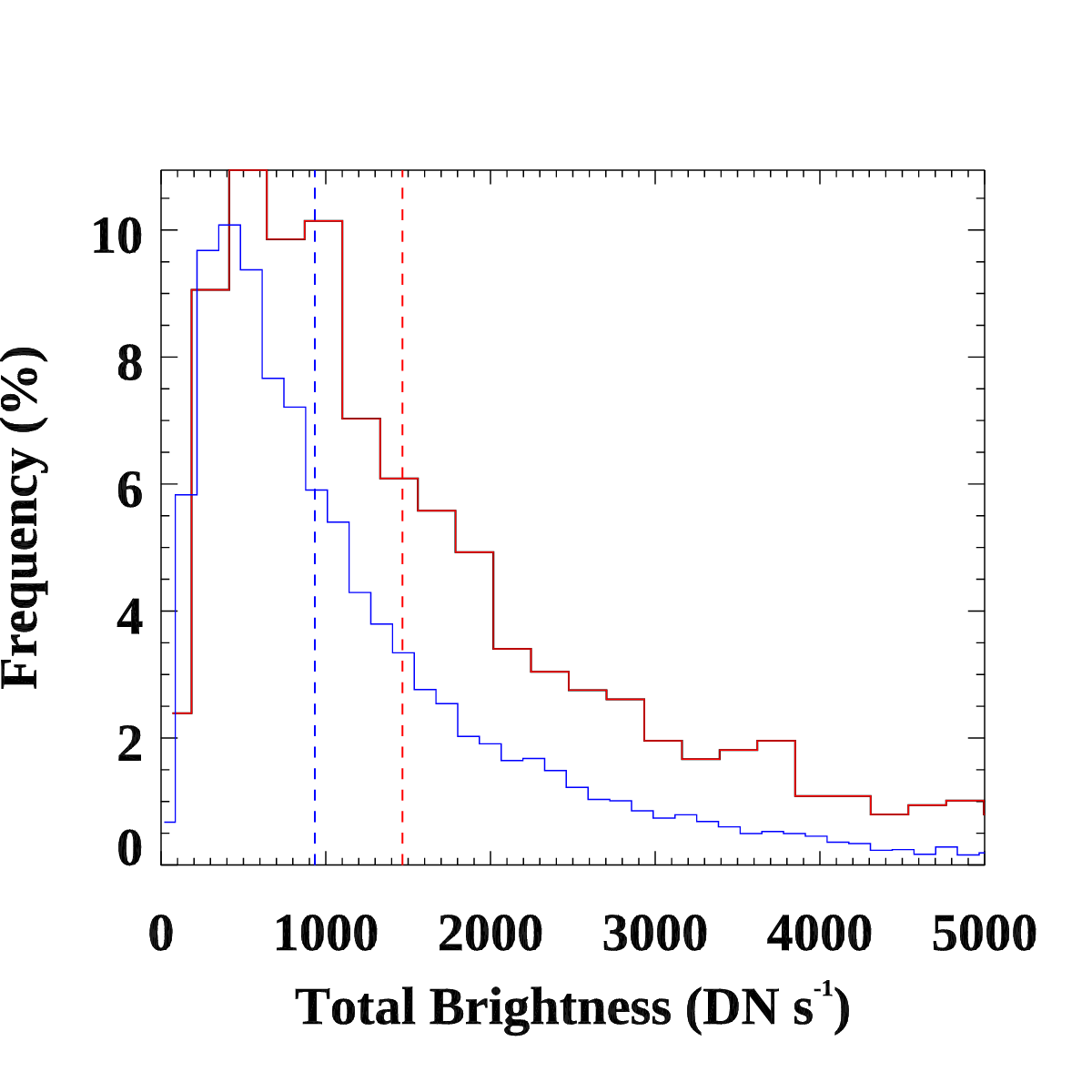}
    (d)\includegraphics[width=0.3\textwidth]{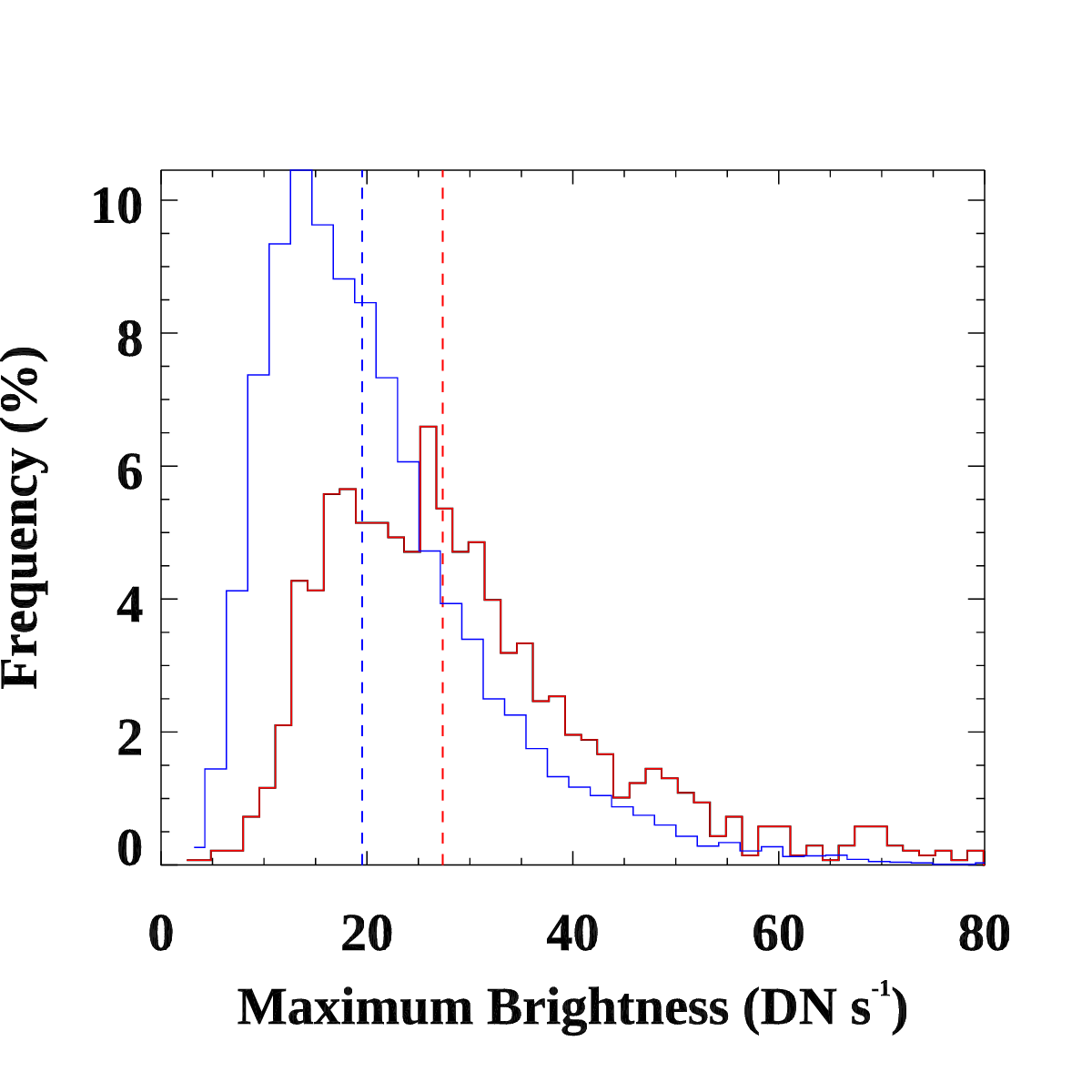}
    (e)\includegraphics[width=0.3\textwidth]{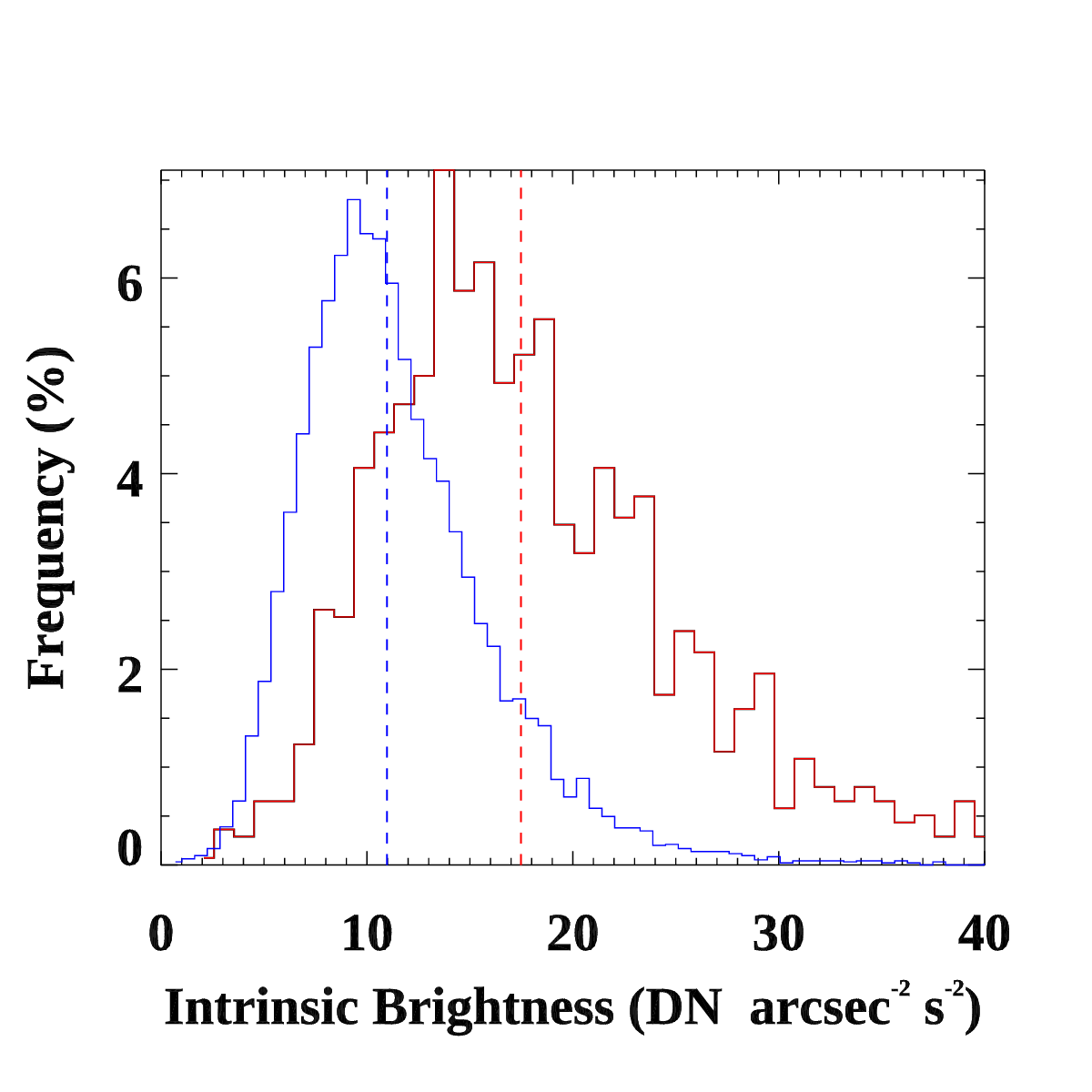}
    (f)\includegraphics[width=0.3\textwidth]{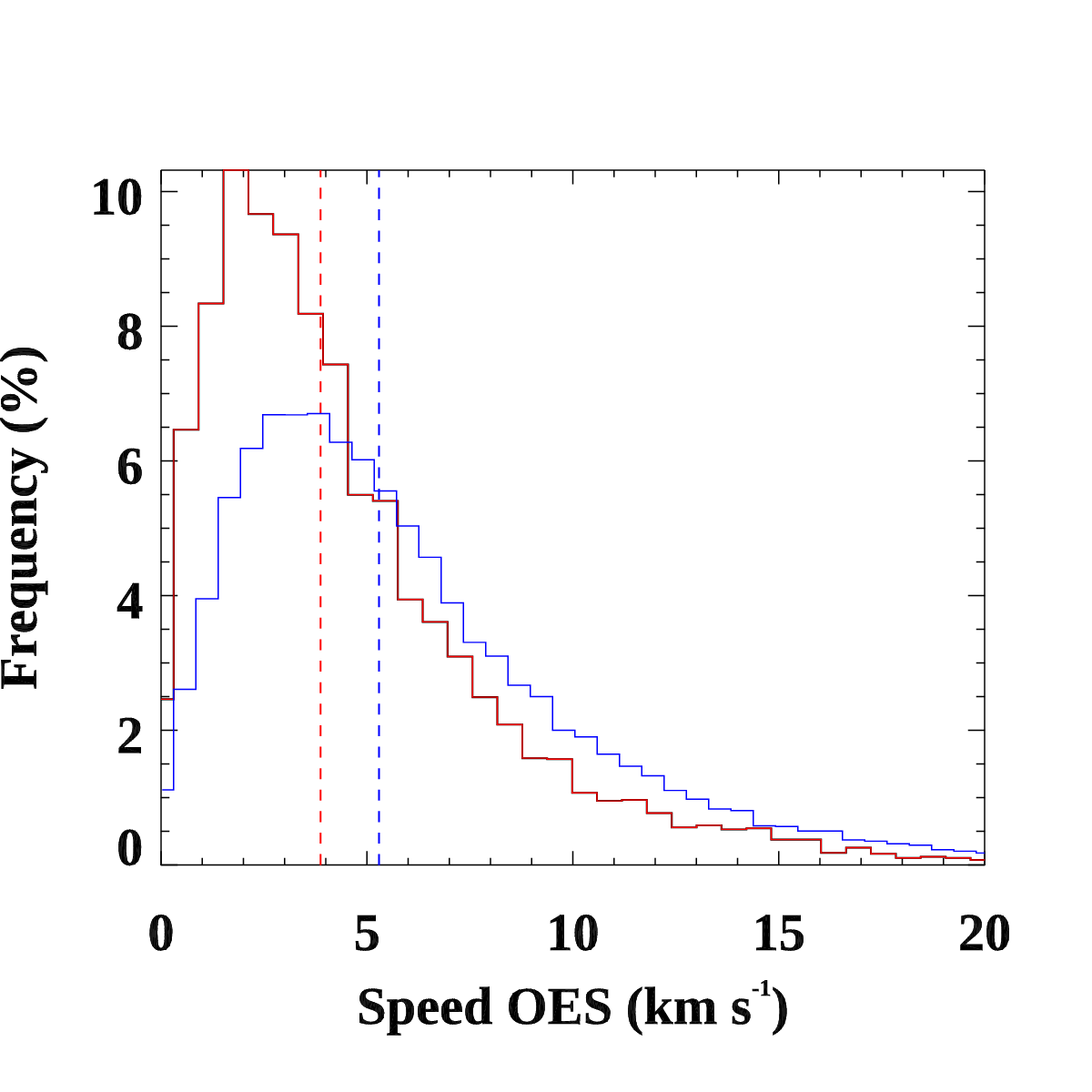}
    (g)\includegraphics[width=0.3\textwidth]{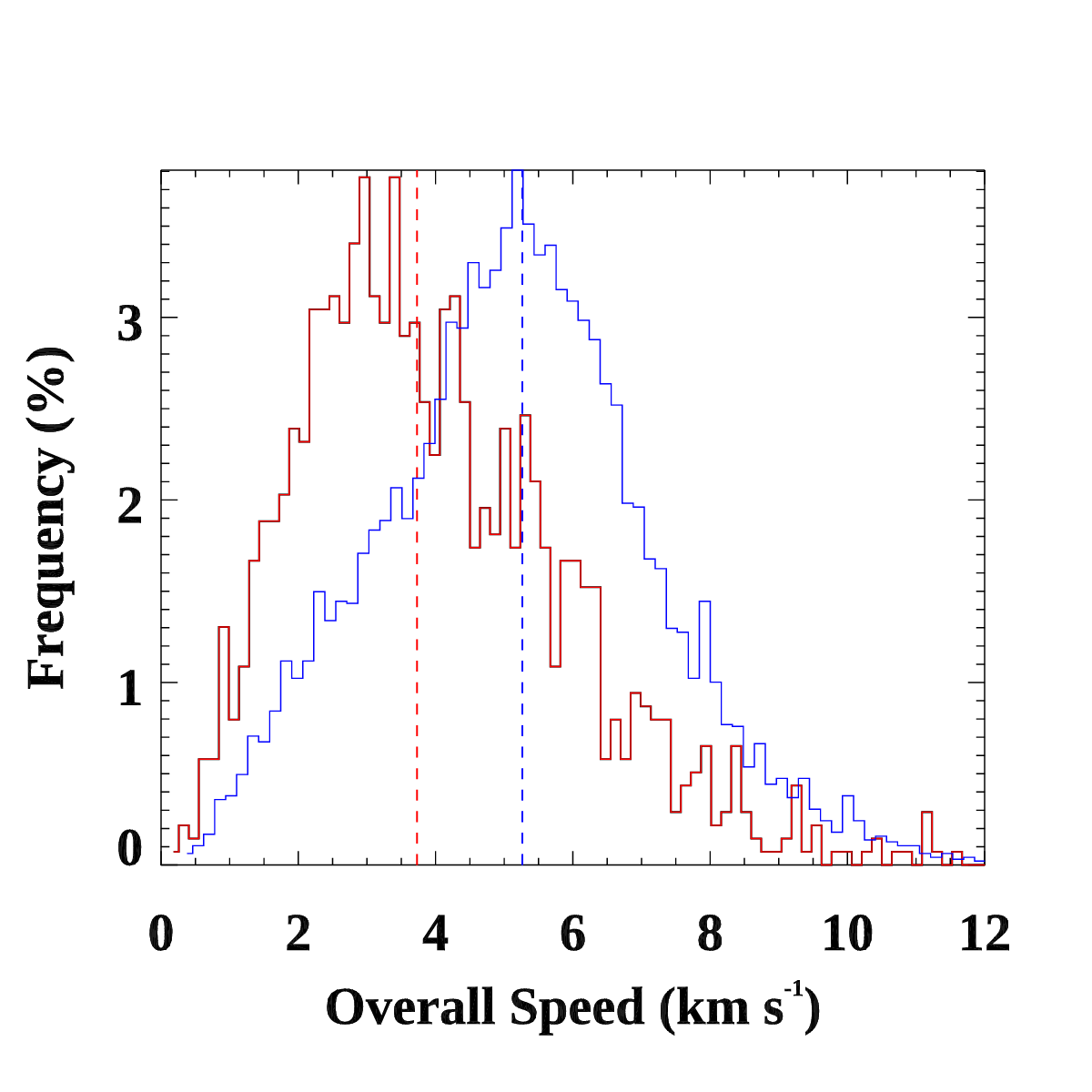}
    (h)\includegraphics[width=0.3\textwidth]{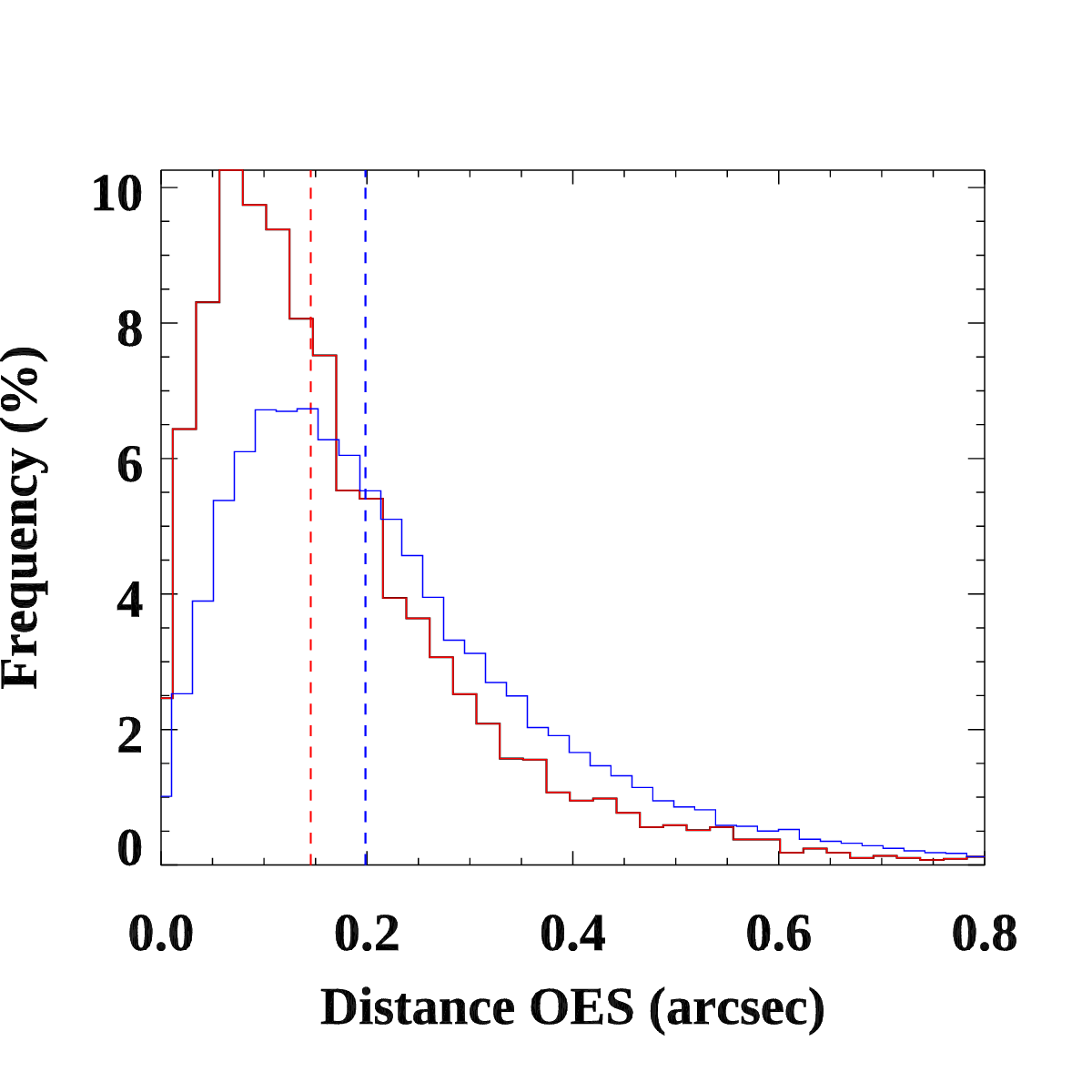}
    (i)\includegraphics[width=0.3\textwidth]{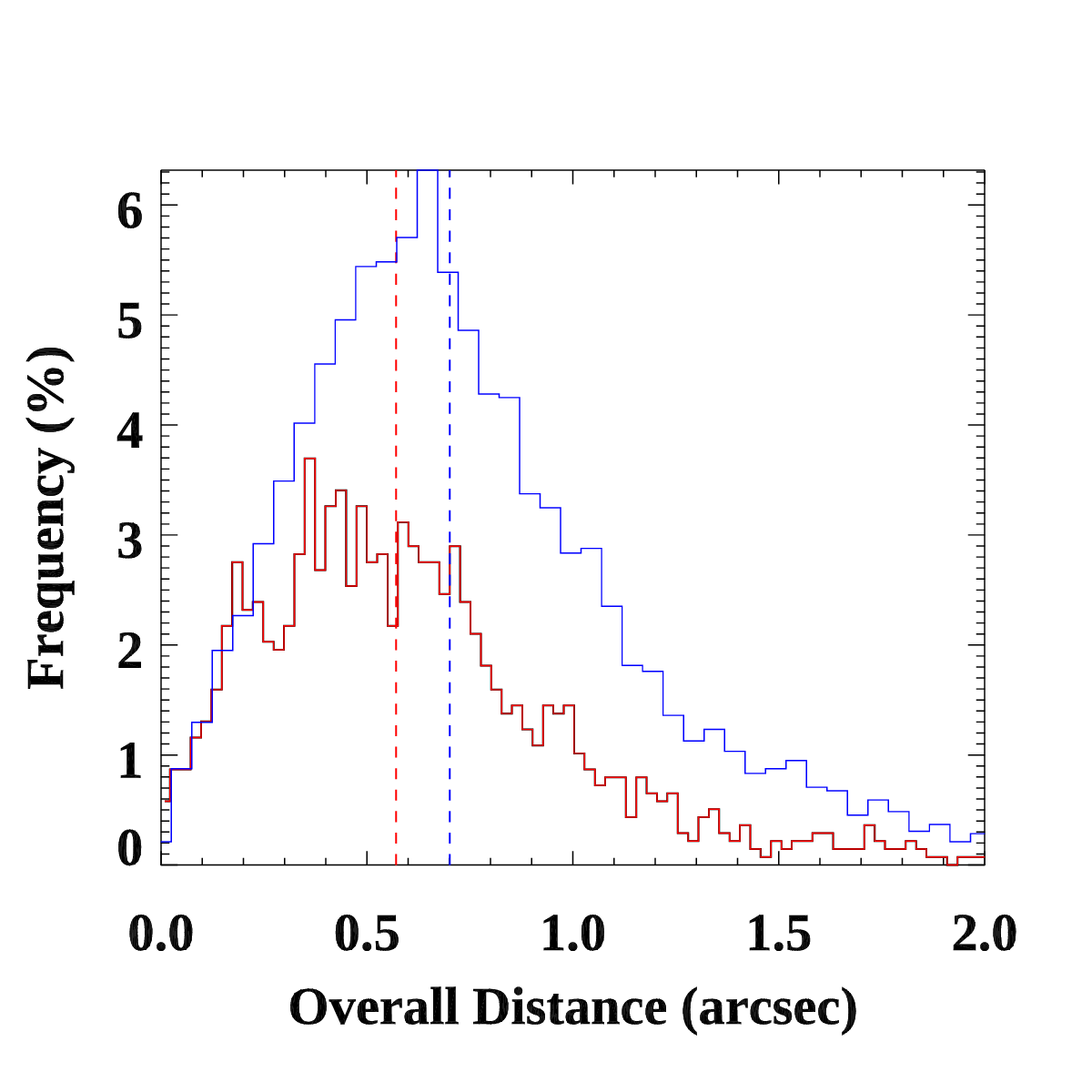}
    (j)\includegraphics[width=0.3\textwidth]{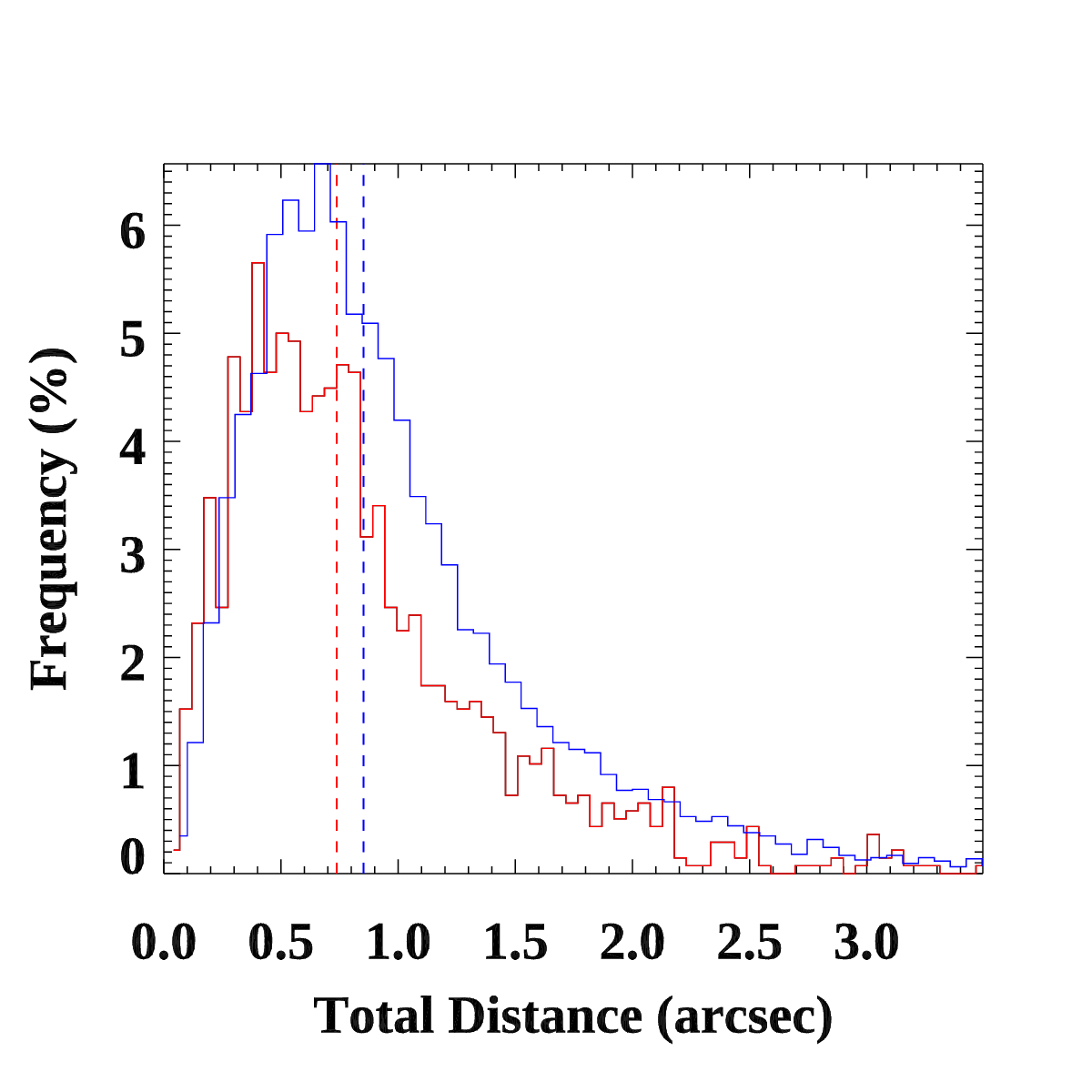}
    (k)\includegraphics[width=0.3\textwidth]{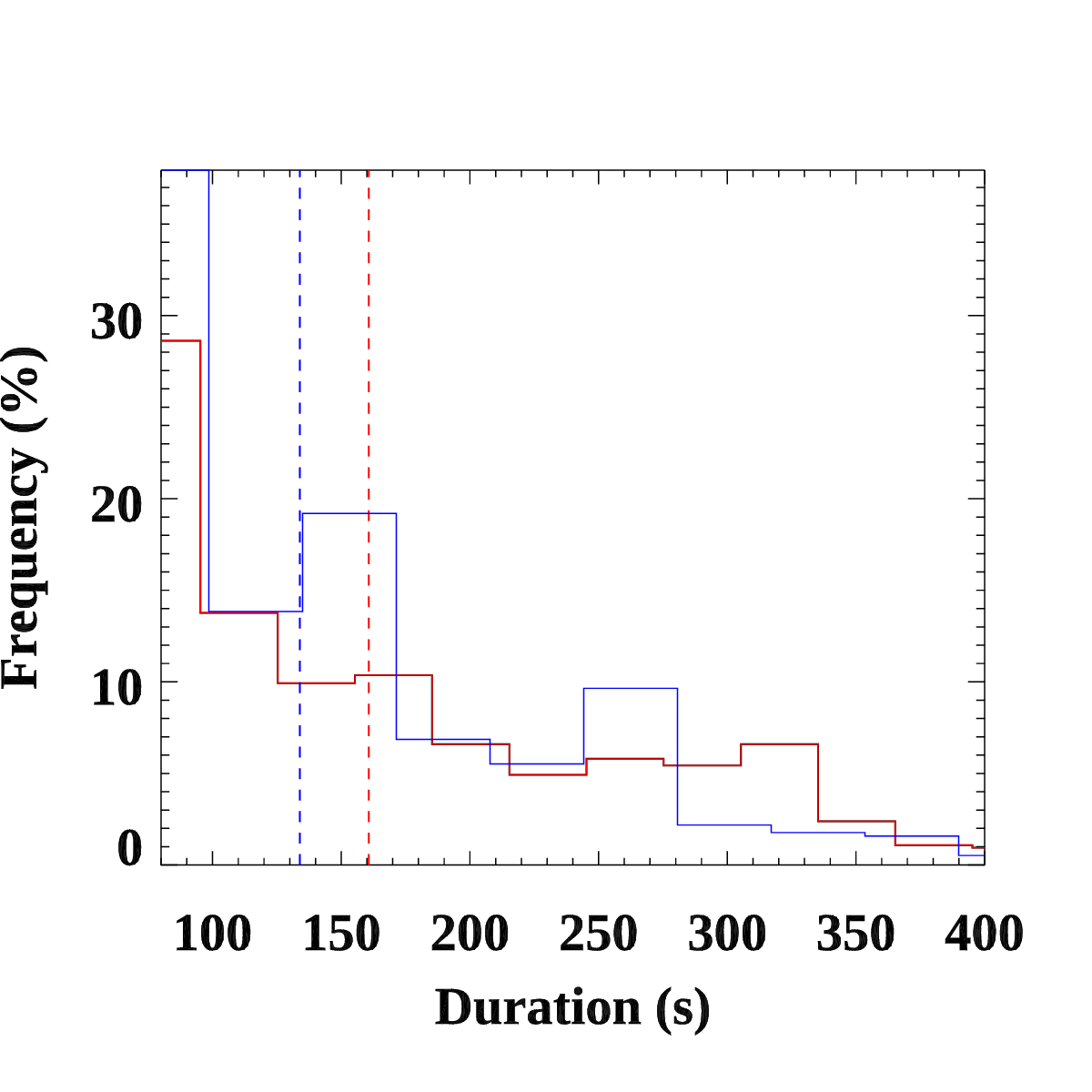}
    \caption{Comparative histograms of 2013-09-26's dataset, whereby red corresponds to AQS and blue corresponds to TQS. Vertical dashed lines represent median values, coloured accordingly. Intrinsic Brightness - defined as the total brightness of an event divided by its area and duration and measured in DN arcsec$^{-2}~$s$^{-2}$ - and Total Distance - defined as the sum of OES distances traversed by a BP - are included.}
    \label{fig:histograms_AQS_vs_TQS}
\end{figure*}

\subsection{Detection Method}\label{sec:sub_method}

The filtering and thresholding procedure is the same as that used in \paperi\ and \paperii. 
Each 1400 \AA\ data cube receives band-pass filtering in space and time by applying three sequential convolutions with vector kernels in the two spatial and one temporal dimension. The bandpass limits the frequency range, whereby frequencies above and below this particular optimal range - 0.09 and 0.40 as determined by \paperi\ - are attenuated such that background noise and slow-changing (and/or large) features do not significantly influence the detection process. A spatially-varying threshold is then defined as a multiple of the standard deviation over time of a filtered analogue synthetic ``noise" cube (see \paperi\ for details on generating synthetic data cubes based on Poisson statistics), together with a second, lower threshold that enlarges the region around each detection (whereby more accurate readings of BP area can be determined). All voxels within these enlarged regions that are above the lower threshold are considered part of their corresponding event detected above the higher threshold (see \paperi\ for details on under-estimation of detected area values). It is worth noting that, whilst the threshold is linked to the region intensity (as a multiple of the standard deviation of the filtered noise), it is based on a Poisson noise model, and while we cannot rule out a selection bias, it is difficult to envisage an alternative approach.

Several characteristics for each detected BP are recorded, such as the average area across a BP's lifetime, overall speed, overall travel distance, and overall Angle of Motion (AOM), each of which is listed in table \ref{tab:data_AQS_vs_TQS}. ``Overall" here denotes the difference between the start point and
end point of a BP, and AOM is defined as the angle at which a BP moves relative to the vertical plane of the FOV. In addition to these average/overall parameters, their values are also recorded for each observation over a BP's recorded lifetime.
%BP's `step', i.e., each frame in which a BP is recorded. 
These are denoted as ``Of Each Step" (OES) results and are recorded along side their overall results in table \ref{tab:data_AQS_vs_TQS}. Duration, maximum brightness and total brightness of each BP are also recorded. An additional parameter - intrinsic brightness - defined as the total brightness divided by both the duration and mean area, while not included in table \ref{tab:data_AQS_vs_TQS}, is also recorded and will be discussed shortly. It is worth noting that these brightness values are background subtracted, whereby the background intensity is determined by taking the median value of the surrounding boxed area. Based on the determined area of the BP, we define a box where the x- and y-range are extended by a handful of pixels relative to a BP’s minimum and maximum spatial coordinates. The intensities of pixels within the box but not within the BP itself are recorded and the background intensity is calculated as the median value.

The event density and \nfrag\ parameters are also recorded; event density describes the number of events detected within a particular FOV within a particular datacube's duration ($\rm arcsec^{-2}~s^{-1}$), and the \nfrag\ parameter describes the greatest number of isolated regions of an event during its lifetime (see \paperi\ or \paperii\ for details). For this study, we focus solely on detections which either do not fragment \nfrag$=1$ or fragment only once \nfrag$=2$ as to avoid overly complex BPs. The number of each dataset's detections that are appropriately simplistic, as well as the percentage of the total number of brightenings that fit this criterion, are listed in table \ref{tab:data_AQS_vs_TQS}.

Candidate brightenings that are smaller than 25 voxels are discarded. A 3-frame minimum BP duration is chosen for this study over \paperii's 5-frame minimum duration due to a difference in cadence ($\sim27$s and $\sim17$s, respectively), equating to approximately a similar $\sim85s$ lower cut-off point for an event's duration. This 3-frame limit will reduce the likelihood of any false positive detections. However, users of the code - readily available on \href{https://github.com/llyrh/Chromosphere_detection_code}{Github} - may chose whichever parameters they see fit for detecting BPs.

There are no upper limits to area or duration. There is an intrinsic upper limit to an event's motion: if an event is moving so rapidly that there is no spatial overlap between one frame and a subsequent frame, then these events will be initially detected as separate events and subsequently discarded due to the 3-frame lower duration limit. Provided that a detection's intensity is above the Poisson-noise-based threshold, there are no fixed upper/lower limits on intensity (see section 2.2 of \paperii\ for details).

\begin{figure*}
    \centering
    (a)\includegraphics[width=0.3\textwidth]{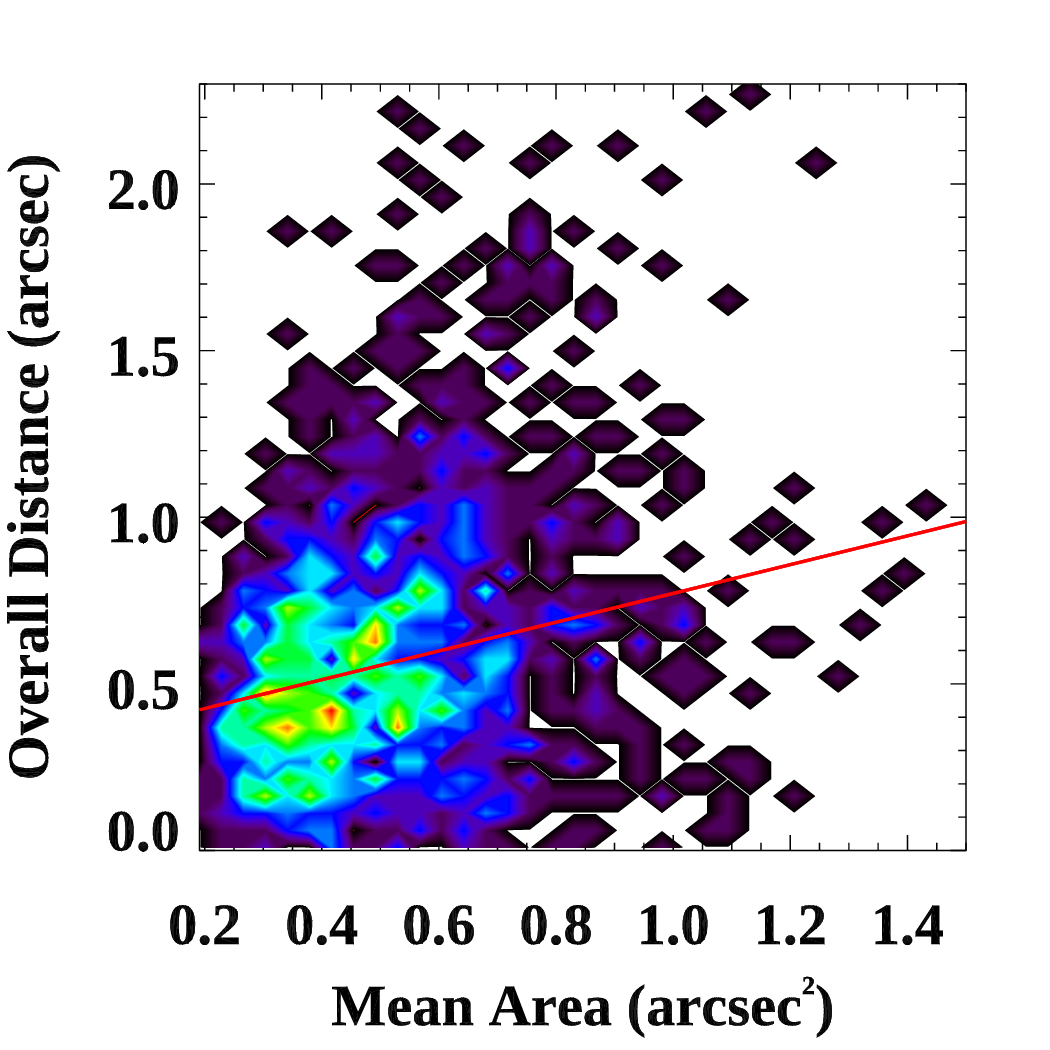}
    (b)\includegraphics[width=0.3\textwidth]{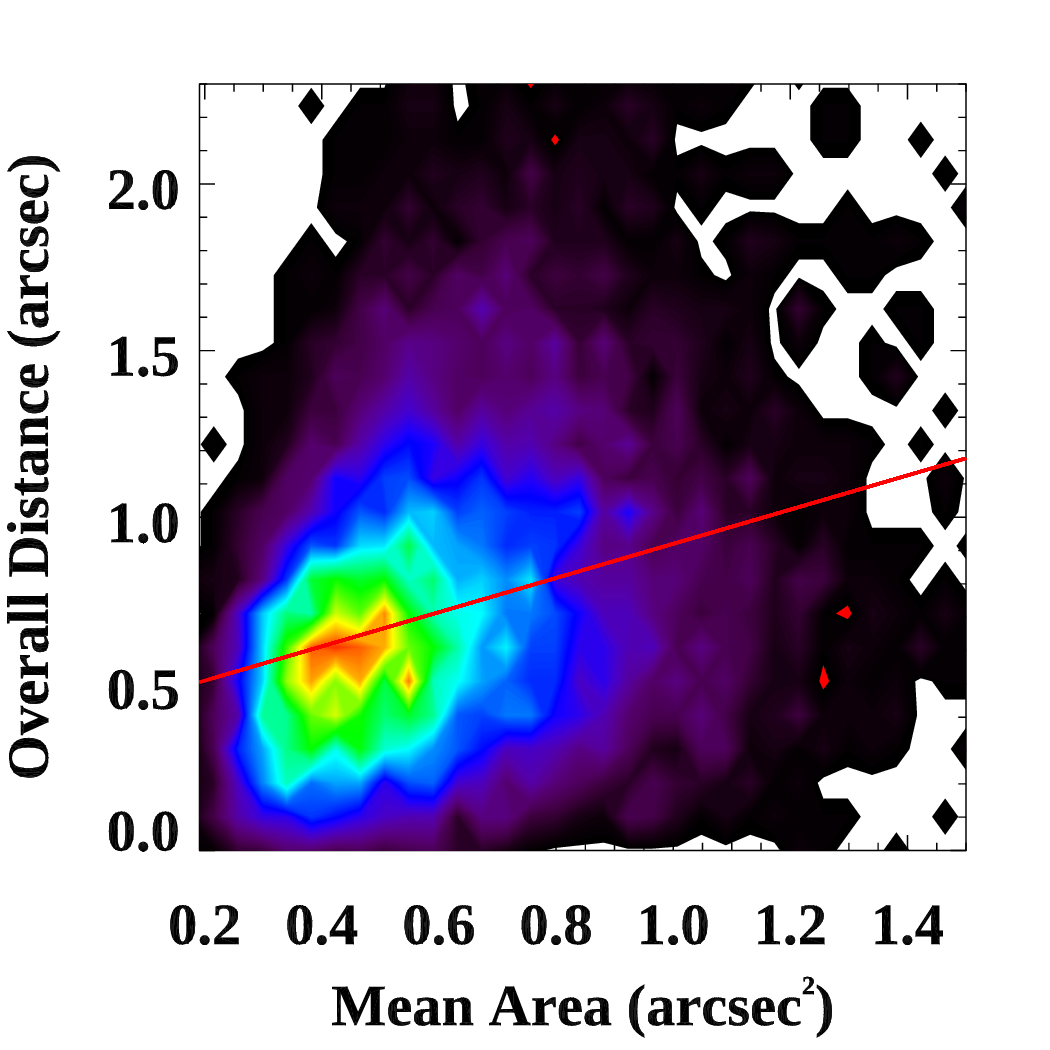}\\
    (c)\includegraphics[width=0.3\textwidth]{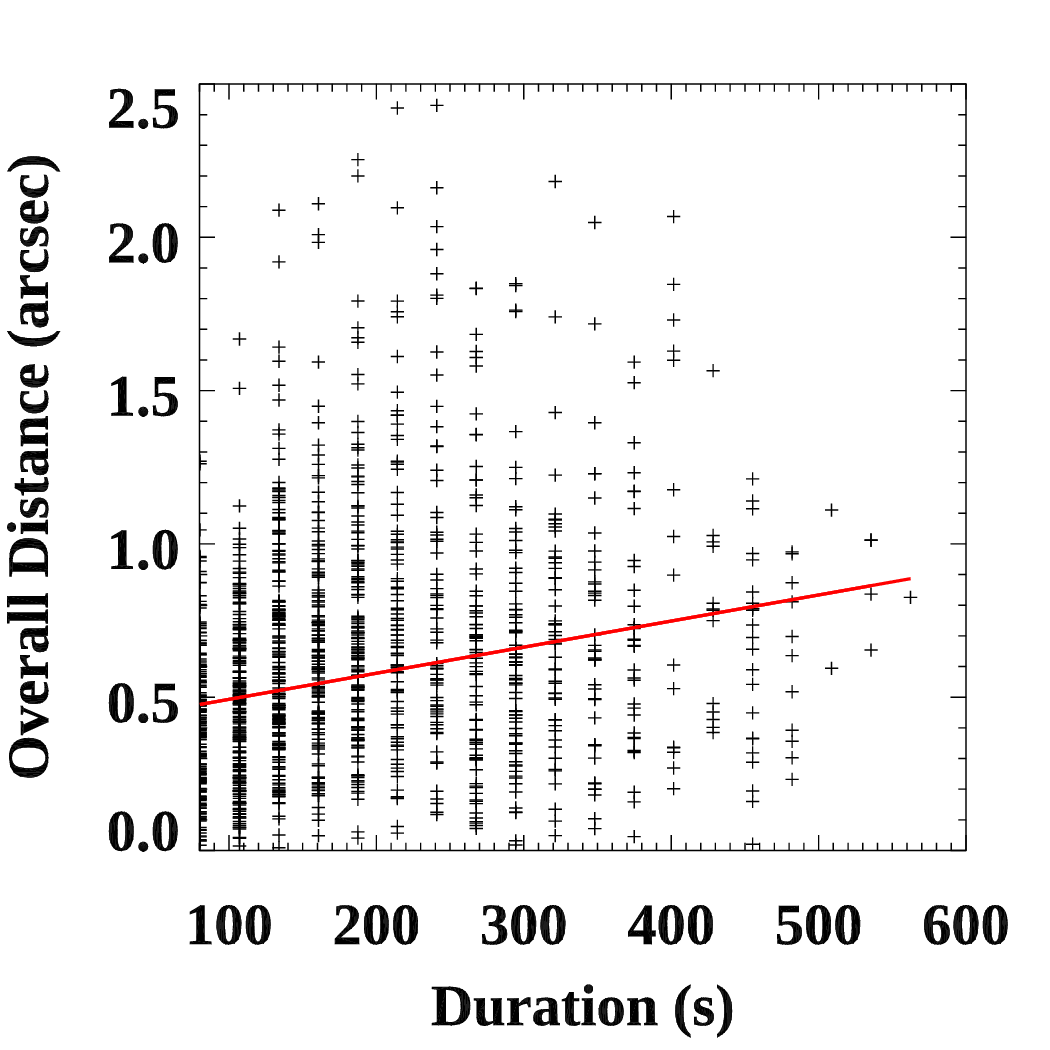}
    (d)\includegraphics[width=0.3\textwidth]{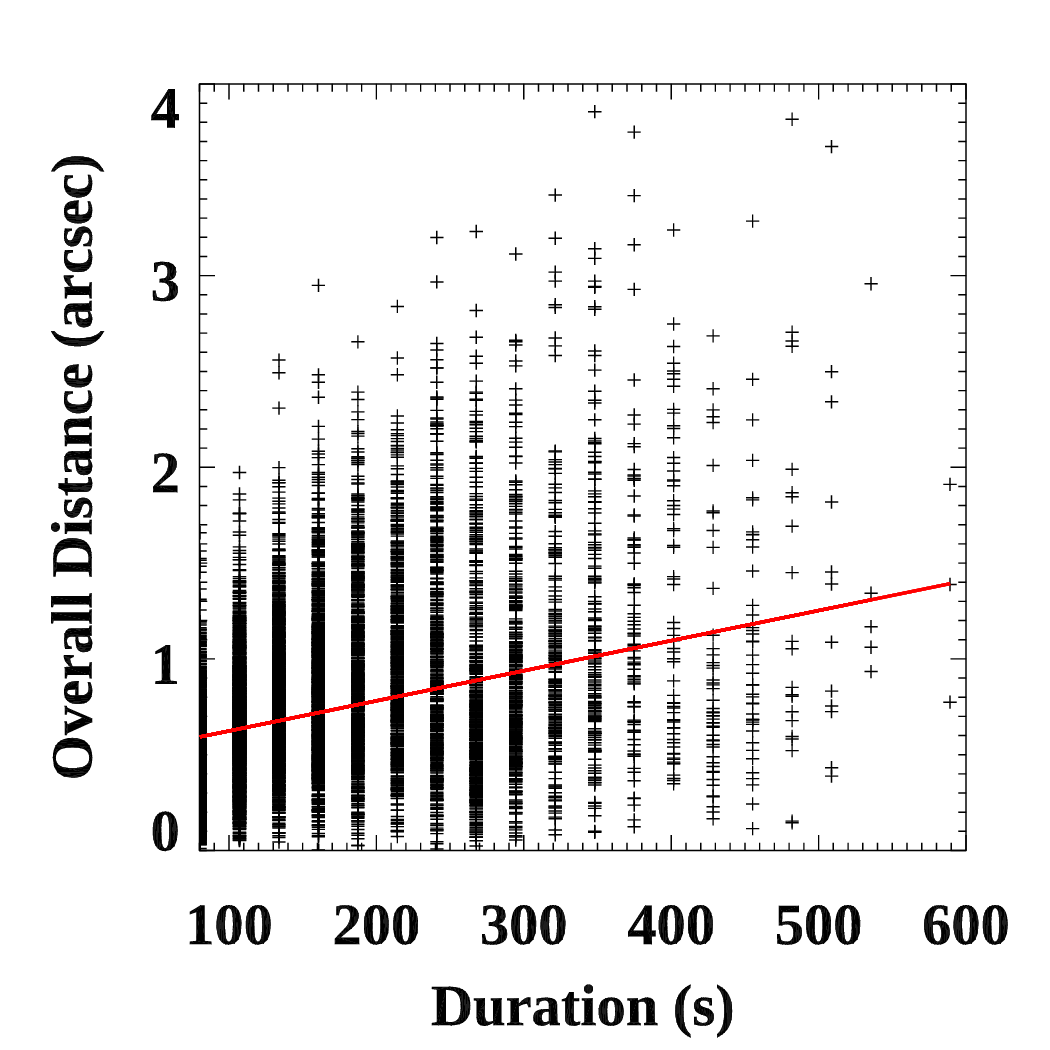}\\
    (e)\includegraphics[width=0.3\textwidth]{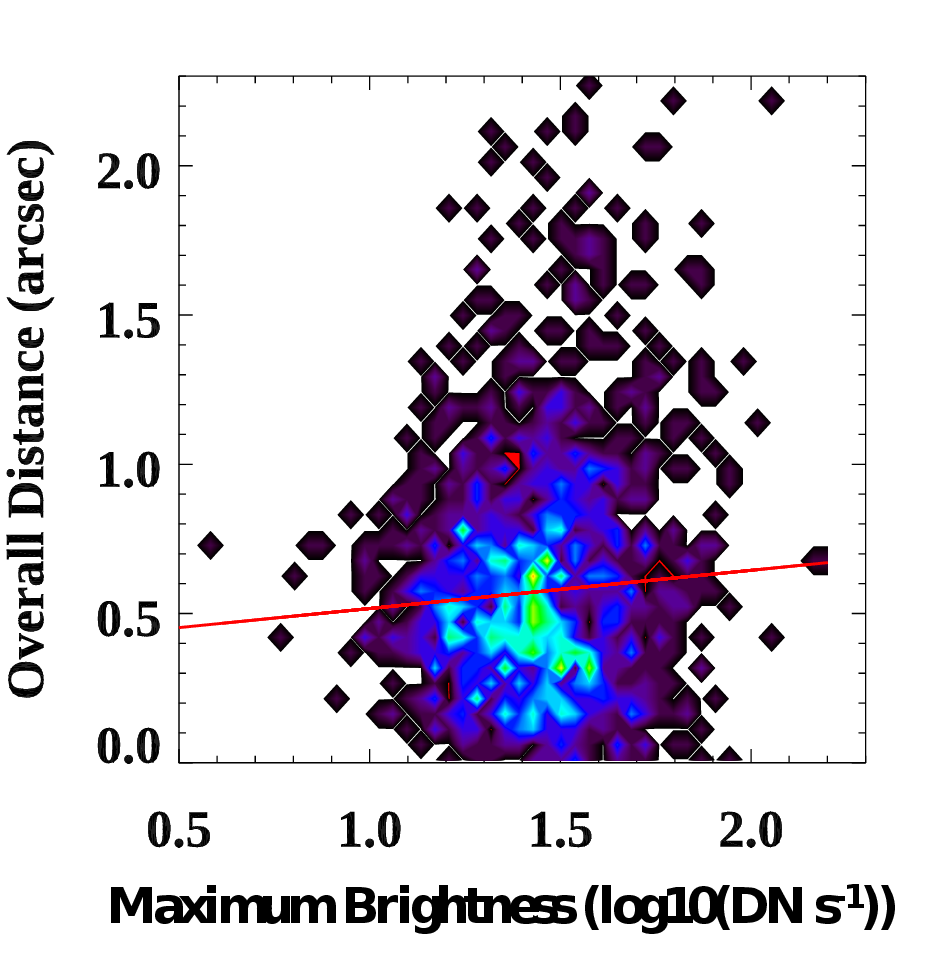}
    (f)\includegraphics[width=0.3\textwidth]{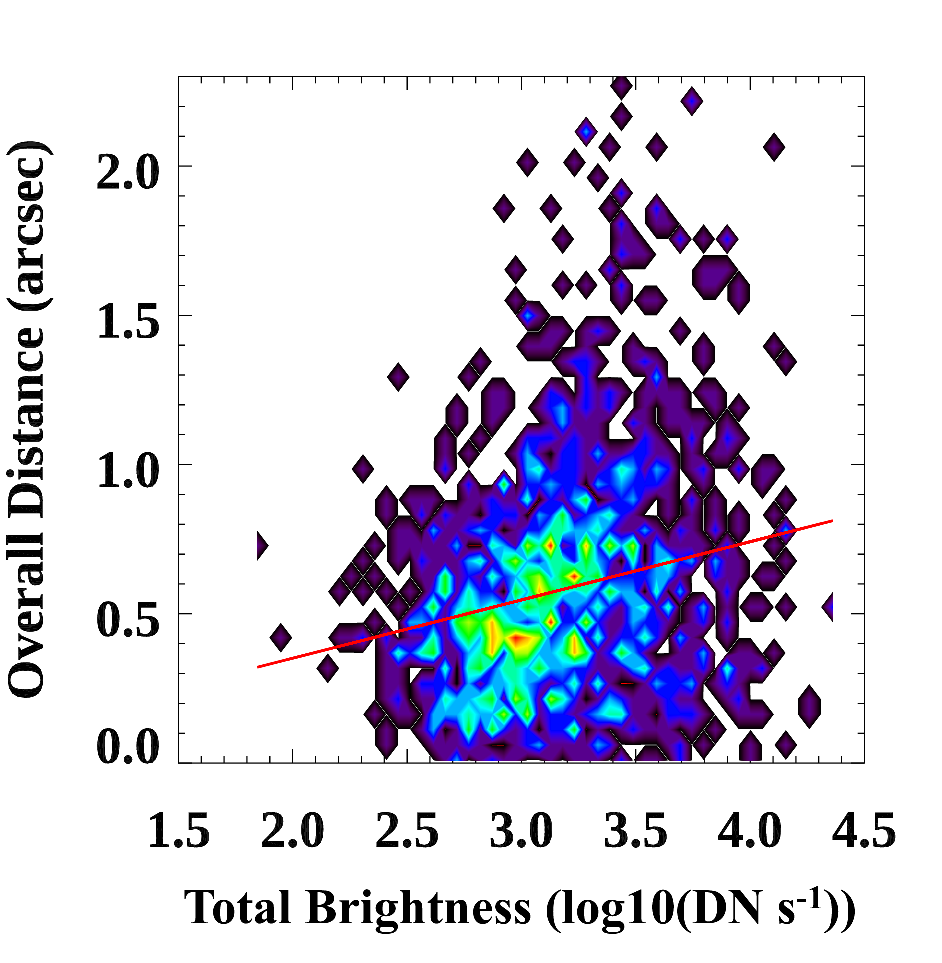}
    (g)\includegraphics[width=0.3\textwidth]{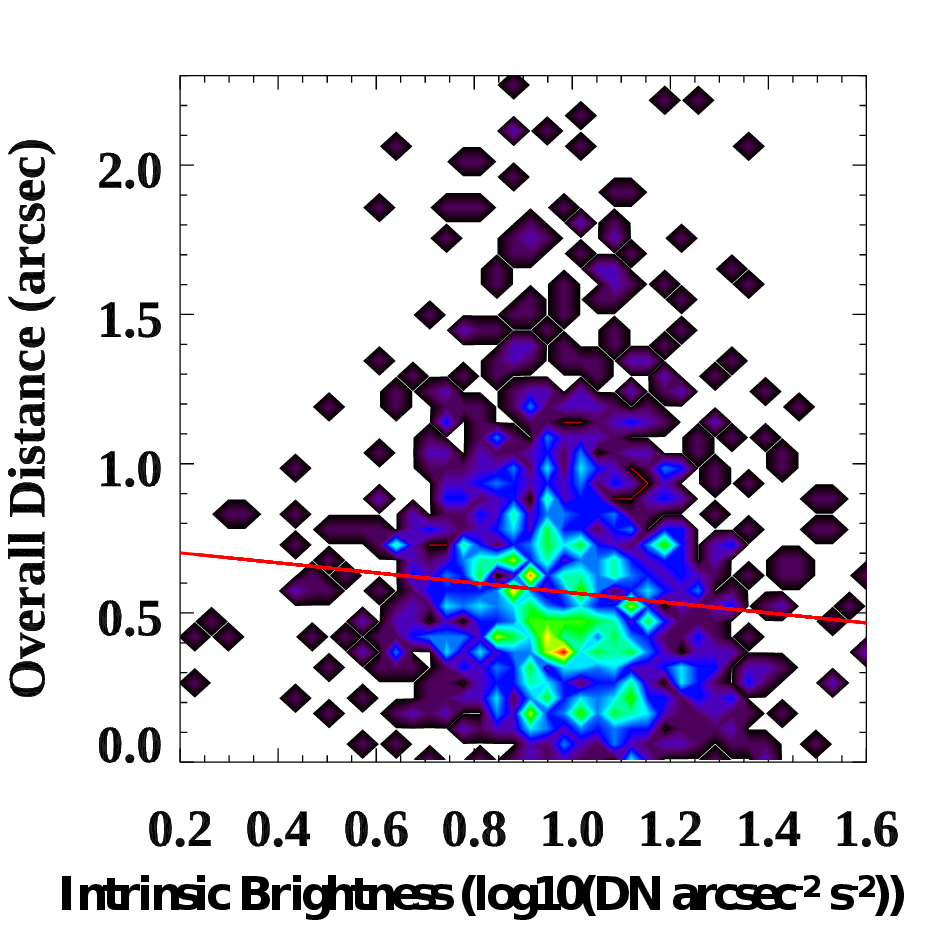}
    (h)\includegraphics[width=0.3\textwidth]{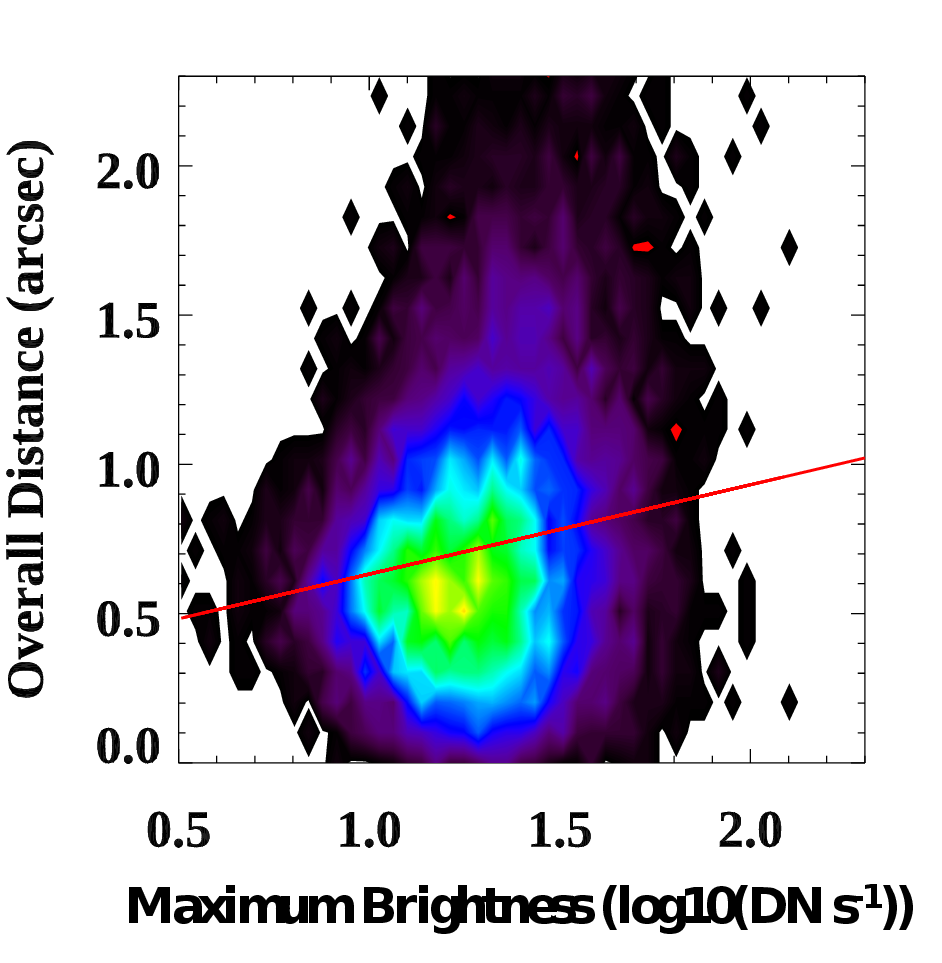}
    (i)\includegraphics[width=0.3\textwidth]{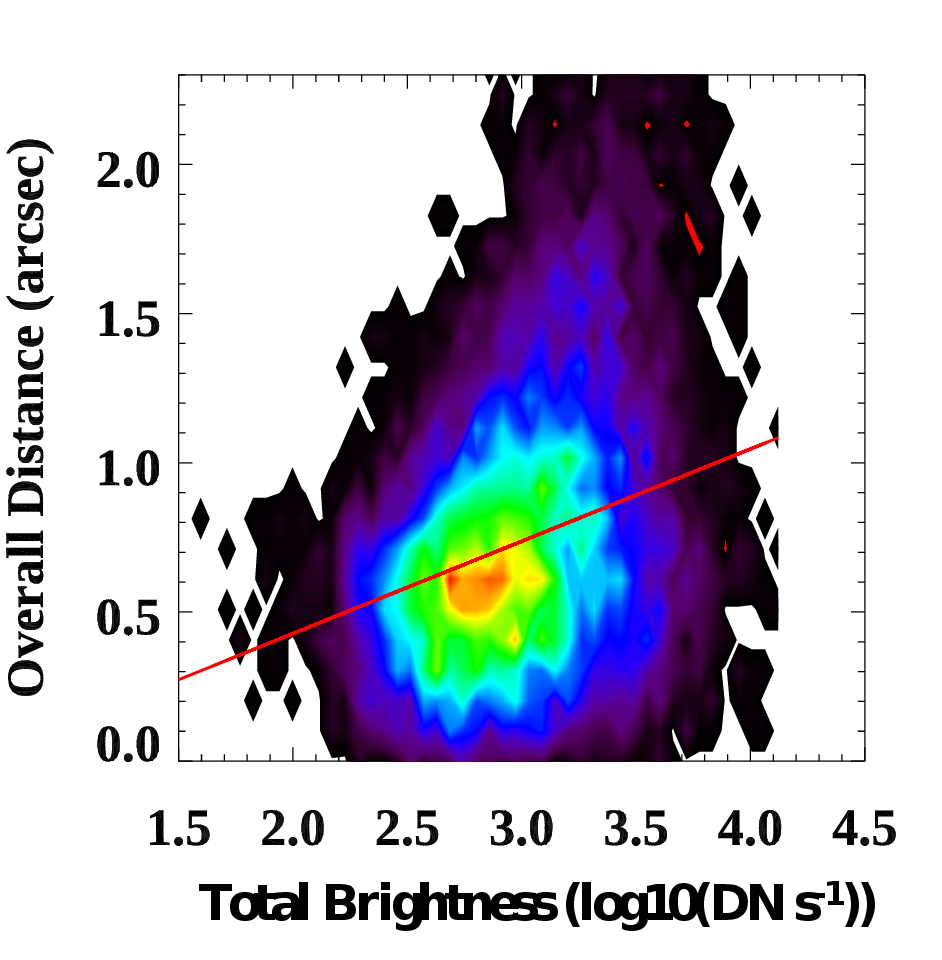}
    (j)\includegraphics[width=0.3\textwidth]{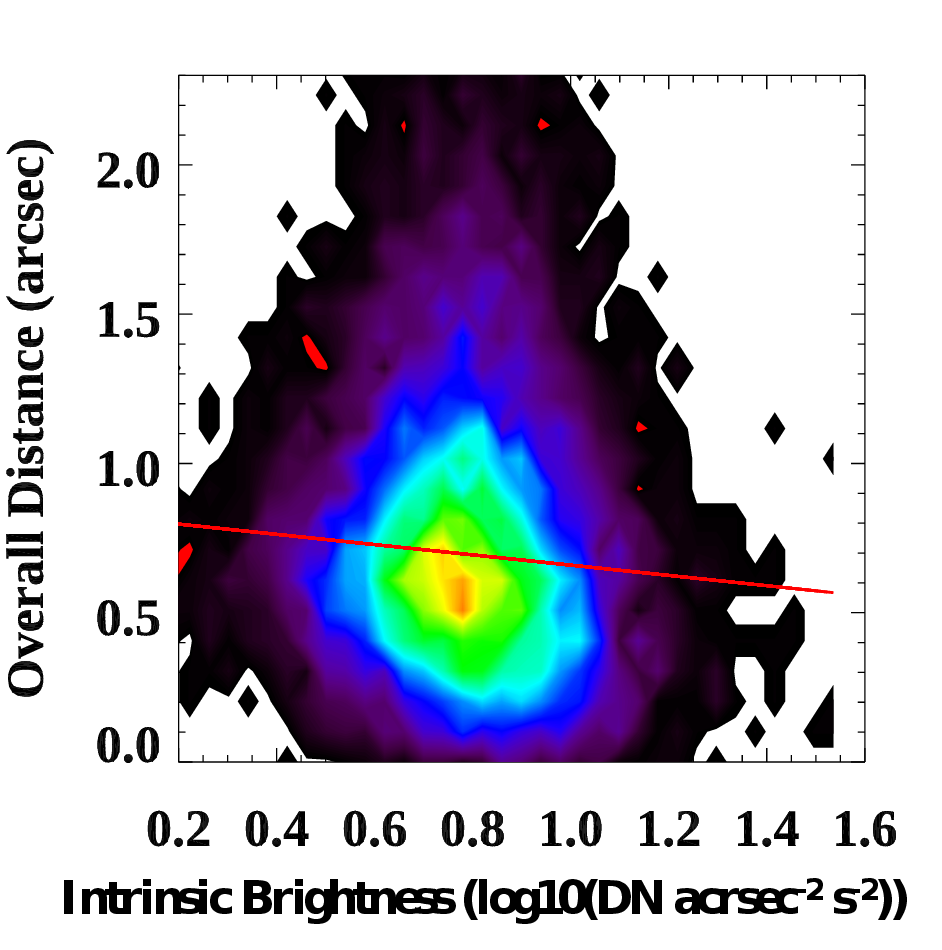}
    \caption{2D Histogram analysis of BP brightness vs overall distance. Each plot has overall distance on the y-axis: (a) AQS mean area, (b) TQS mean area, (c) AQS duration, (d) TQS duration. The remaining plots are of detected brightness values: (e) AQS maximum brightness, (f) AQS total brightness, (g) AQS intrinsic brightness, (h) TQS maximum brightness, (i) TQS total brightness, (j) TQS intrinsic brightness. Red lines are the result of least absolute deviation fits.}
    \label{fig:2D-hist}
\end{figure*}

\subsection{``Active" vs ``True" Quiet-Sun}

Masking and thresholding of the IRIS 1400 \AA\ data defines observations for relatively bright and dim domains, which we donate as ``Active" Quiet-Sun (AQS) and ``True" Quiet-Sun (TQS), respectively and can be seen in figure \ref{fig:FOV_w_contours_and_box}. These limits are defined by first generating a time-averaged image of the 1400 \AA\ dataset (as is displayed in figure \ref{fig:FOV_comp}a) and then ascertaining the mean value of this time-averaged 1400 \AA\ image. AQS domains are defined as $\ge1.4$ times the mean and lie \textit{inside} the red contours, while TQS are defined as $\le1.2$ times the mean and lie \textit{outside} the blue contours. If the mean spatial position of BPs are detected within either of these domains, then they are recorded as either AQS BPs or TQS BPs accordingly. Regions that lie between these contours (as well as any BPs detected therein) are ignored. This is done to avoid ambiguity when defining a hard threshold limit; AQS BPs can be redefined as TQS BPs (and vice versa) if a single threshold is used and then redefined, and some BPs that are first detected in either domain may have mean positions in the opposing domain. 

\begin{figure*}[t]
    \centering
    \includegraphics[trim={0cm 0cm 0cm 8cm},width=0.9\textwidth]{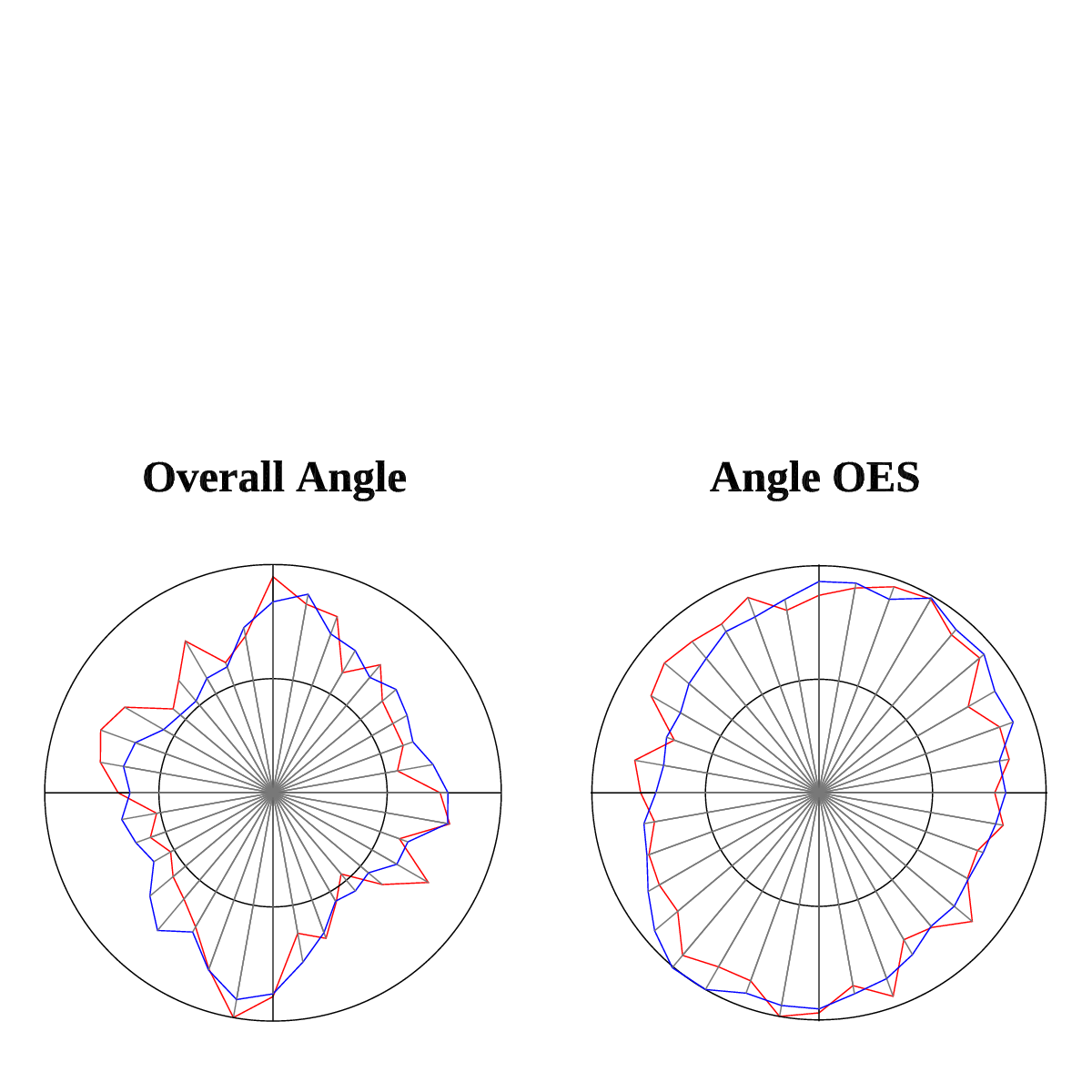}
    \caption{Polar histograms of the overall angle of motion (left) and the angle of motion OES (right). Red indicates AQS. Blue indicates TQS. Results have been normalised.}
    \label{fig:polar_plots}
\end{figure*}

\begin{figure*}[t]
    \centering
    \includegraphics[trim={0cm 0cm 0cm 1.1cm},clip,width=0.8\textwidth]{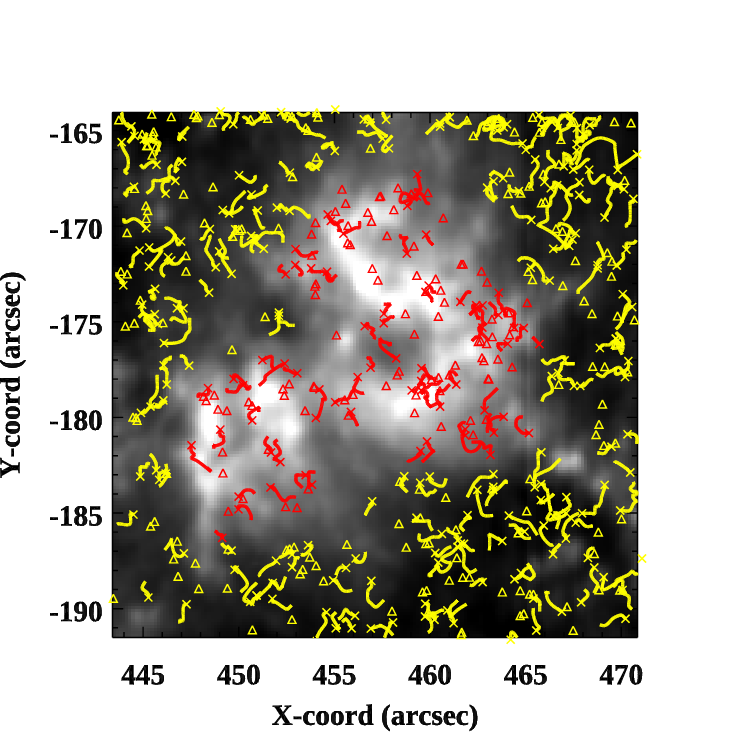}
    \caption{Image of $N_{frag}=1$ (red) AQS and (yellow) TQS BP interpolated motion paths within green box from figure \ref{fig:FOV_w_contours_and_box}. Xs indicate point at which a BP is no longer detectable. Triangles indicate mean location of BPs with very low standard deviations of motion (see text for details).}
    \label{fig:motion_AQS_vs_TQS}
\end{figure*}

\section{Results} \label{sec:results}

\begin{figure*}[t]
    \centering
    \includegraphics[trim={3.5cm 0cm 0cm 0cm},width=0.5\textwidth]{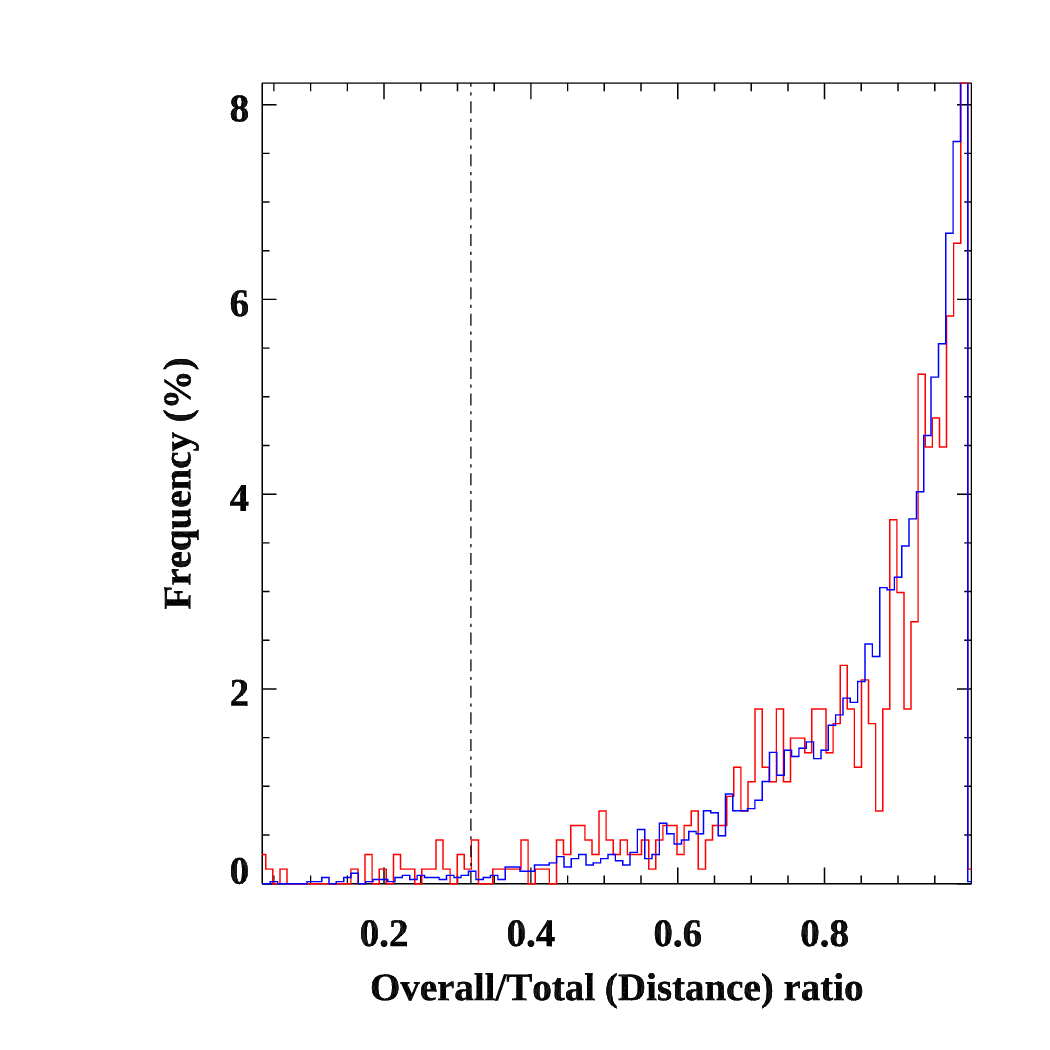}
    \caption{Histogram plot of ratios of overall travel distance to total travel distance. Red correspond to AQS results. Blue corresponds to TQS results. Vertical dashed line is equal to $1/\pi$.}
    \label{fig:ratio}
\end{figure*}

\subsection{AQS/TQS Comparison}

Several BP characteristics are listed in table \ref{tab:data_AQS_vs_TQS}, comparing those within the entire FOV, the AQS domains, and the TQS domains. These statistics are further detailed in figure \ref{fig:histograms_AQS_vs_TQS} which displays red and blue histogram plots for AQS and TQS distributions, respectively, of various properties.
By considering both the data from table \ref{tab:data_AQS_vs_TQS} and figure \ref{fig:histograms_AQS_vs_TQS}, the following conclusions can be made:
\begin{itemize}
    \item TQS BPs tend to be larger than those in AQS domains, whether it be their mean area over their lifetimes, or their area OES
    \item AQS BPs tend to be brighter than their TQS counterparts in all aspects: maximum brightness, total brightness, and intrinsic brightness. Note: while the domains are split based on mean intensity, it does not necessarily follow that BPs from different domains differ in brightness as brightness values are determined after subtracting background intensities
    \item AQS BPs tend to travel at slower speeds and over shorter plane-of sky (POS) distances than those in TQS domains, whether that be their overall travel distance/speed, distances travelled or speeds OES, or their total travel distance
    \item The distribution of AQS and TQS durations are very similar, but AQS BPs tend to last longer than those found in TQS domains
    \item The BP area OES have greater maximum values than their average area values, which suggests that the BP size tends to change during their lifetime. This conclusion agrees with the preliminary findings of \paperi. A closer analysis of the trends in changing area values will be conducted in a future study
    \item Despite the AQS BPs's slightly smaller median area and larger median duration, they are intrinsically brighter, suggesting that AQS BPs posses greater free magnetic energy
\end{itemize}

Note that while the conclusions above are based on our rigorous detection criteria, we cannot necessarily rule out a selection bias based on active domains being intrinsically brighter than quiet domains. The following subsections will discuss some of these aspects in greater detail.

\subsection{Motions Analysis}\label{sec:motions}

Figure \ref{fig:2D-hist} presents an analysis of BP brightness values and their overall travel distances. The top row demonstrates mean area vs overall distance for (a) AQS and (b) TQS. The second row plots duration vs overall distance for (c) AQS and (d) TQS results. The third row (e, f and g) represents AQS domain results of brightness vs overall distance, while the bottom row (h, i and j) represents those from TQS domains. The red lines are the result of a robust least absolute deviation (LAD) fit method. (a) and (b)'s 2D histograms show that there is in fact a relationship between mean area and overall distance, with LAD gradients of 0.43 and 0.51 for AQS and TQS, respectively. Fig (c) and (d) have LAD gradients of 0.0008 and 0.0016, respectively. While these gradients suggest no correlation, this may be due to duration values being restricted to integer multiples of the mean dataset cadence. Incidentally, this is also why 2D histograms are unavailable for these plots. These duration and/or size biases are absent from (g) and (j)'s intrinsic brightness plots.

Maximum brightness (e and h) and total brightness (f and i) show a slight tendency for the brightest events to travel the greatest POS distances, with LAD gradients of 0.13, 0.20, 0.30, and 0.31, respectively. The larger LAD gradients for TQS BPs suggest a stronger correlation between brightness and overall travel distance in these domains. If greater brightness values tend to correlate with greater travel distances, then it would be expected that the AQS LAD gradient would be greater than that of TQS (as figure \ref{fig:histograms_AQS_vs_TQS} and table \ref{tab:data_AQS_vs_TQS} have shown that AQS BPs tend to be brighter), but this is not the case. The relationship appears greater for total brightness, but total brightness has an inherent bias of the duration and/or area of BPs, and intuitively, longer-lived BPs would likely travel further distances. Therefore, the relationship between total brightness and overall distance is unsurprising.

Both intrinsic brightness results for (g) AQS and (j) TQS domains demonstrate LAD gradients of $\sim-0.17$, suggesting a slight inverse correlation with overall travel distance. This will be discussed further in section \ref{sec:disc}.

The overall AOM and AOM OES results in table \ref{tab:data_AQS_vs_TQS} are an attempt to determine any preferential direction of motion of these BPs.
Figure \ref{fig:polar_plots} shows polar histograms of BP motion directions ranging from -180$^{\circ}$ to 180$^{\circ}$, with each segment binned to 10$^{\circ}$. Red and Blue represent AQS and TQS results, respectively. The overall angle - the angle of motion between a BP's start and end points - is displayed on the left, while the angle OES - the angle of motion between a BP's centroid position and a subsequent position across its lifetime - is displayed on the right. There is a slight North-South preference in overall angle (as is also suggested by \paperii). AQS and TQS BP motions are generally similar. There are a large number of detections (over 12,600) and these motions are highly varied and follow complex paths.
Although sensitive to noise, \paperii's contemporaneous results show similarities of motions between data from each of IRIS's slit-jaw imaging channels. The horizontal motions of the angle OES plot in figure \ref{fig:polar_plots} are $\sim80\%$ that of the vertical motions - i.e., the motions along the FOV when viewed from above corresponding to North-South and West-East - and this is likely due to projection effects from the FOV's position on the solar surface; the dataset is centered at $X = 468\arcsec$, $Y = -194\arcsec$, which could produce relatively significant projection effects compared to a FOV at disk-centre. This may also explain why the angle OES plot demonstrates a tendency to skew away from motions along North-South and West-East directions, as motions along these directions become increasingly unlikely as a dataset's FOV approaches the solar limb.
We therefore have confidence that statistically, these detected BP motions are representative of true BP motions and that there are consistent patterns in these motions.

The analysis of these BP motions is successively focused on a smaller ROI, namely that which is encapsulated within the green box seen in figure \ref{fig:FOV_w_contours_and_box}, whereby a large, bright AQS domain is present, surrounded by a darker TQS domain. The BP motions within this smaller region are plotted in figure \ref{fig:motion_AQS_vs_TQS}, whereby red represents AQS BPs and yellow represents TQS BPs. Yellow is chosen to represent TQS in this instance as to avoid sharp contrast with the dark background of the IRIS 1400 \AA\ image. For plotting purposes, these motion plots are restricted to \nfrag$=1$ BP detections, as the image becomes too crowded and incoherent otherwise. Xs indicate where BPs, having followed their interpolated paths, are no longer detectable. Triangles indicate detections where their motions have very low standard deviations - this makes it difficult to interpolate their paths of motion with extreme curves and a small number of points with which to plot, and are therefore ignored for plotting purposes. 

It is difficult to determine from figure \ref{fig:motion_AQS_vs_TQS} whether a preferred motion path is present in either domains. Figure \ref{fig:ratio} provides a closer statistical look at these path motions by providing a histogram plot of the ratio between the overall distance - the distance between the start point and end point of a BP's motion - and the total distance, or the sum of the distance travelled by a BP during its lifetime. In order to correspond more closely with figure \ref{fig:motion_AQS_vs_TQS}, these ratio values are also restricted to \nfrag$=1$ BP detections. The (red) AQS and (blue) TQS distributions are largely similar. The majority of BP motions are greater than the $1/\pi$ vertical dashes line and are near a ratio value of 1, i.e., whereby the overall distance and total distance are equal, which indicates that most BPs follow mostly straight motion paths while only a minority of BPs follow an extremely curved path.

These motions provide a great deal of insight into the behaviour of BPs and warrant further analysis in a subsequent study, particularly if these motions appear contemporaneous between IRIS and/or AIA wavelength channels.

\subsection{Spectroscopic Analysis}\label{sec:spec}

\begin{figure*}[t]
    \centering
    \includegraphics[trim={0 2cm 0 5cm},clip,width=0.85\textwidth]{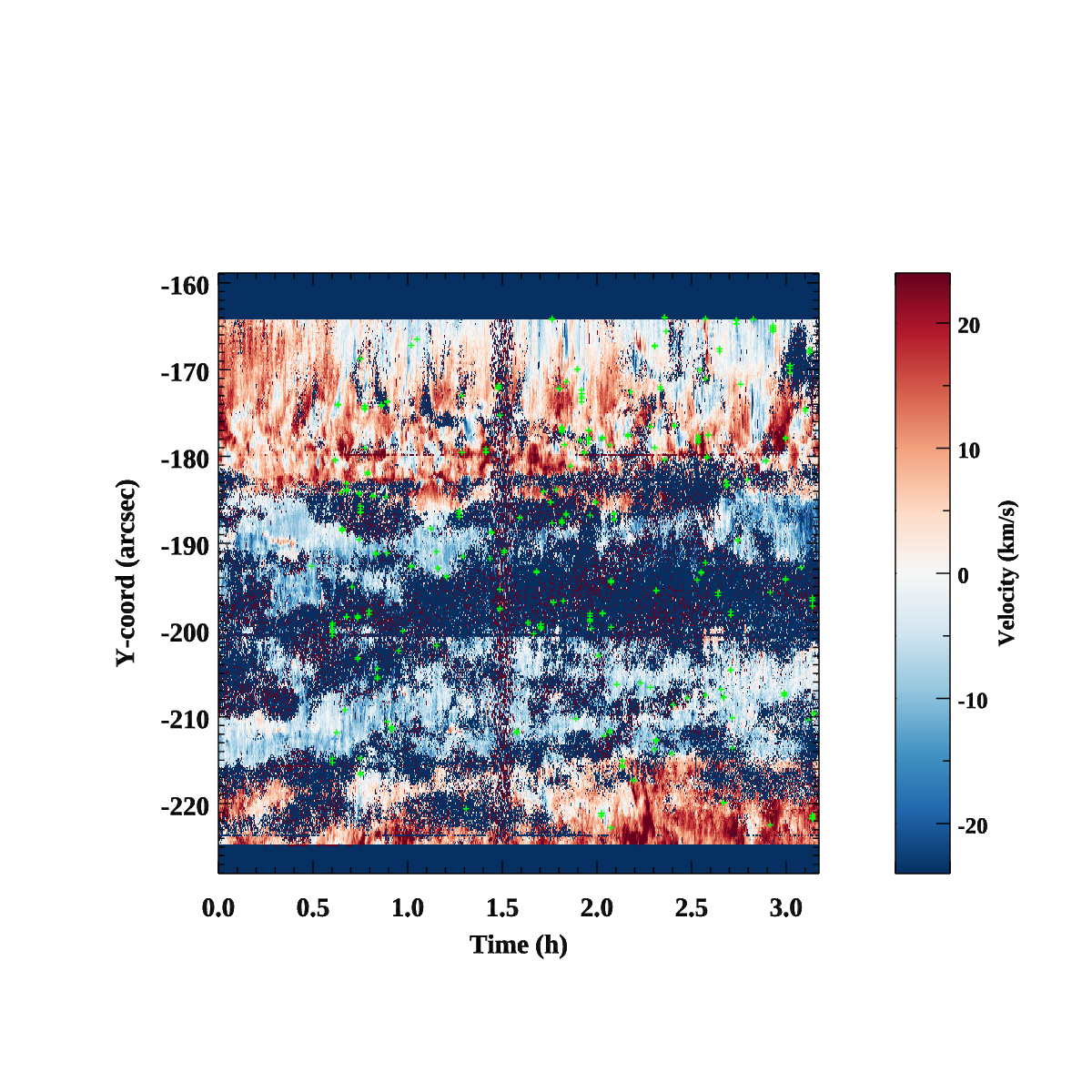}
    \caption{Si IV 1403 Doppler velocity map produced over time as the ROI passes beneath the IRIS slit. Green crosses indicate detected BP positions as they cross the slit position. Color bar and map are limited to a particular velocity range for plotting purposes.}
    \label{fig:Si1403_vel_map}
\end{figure*}

\begin{figure*}[t]
    \centering
    (a)\includegraphics[width=0.40\textwidth]{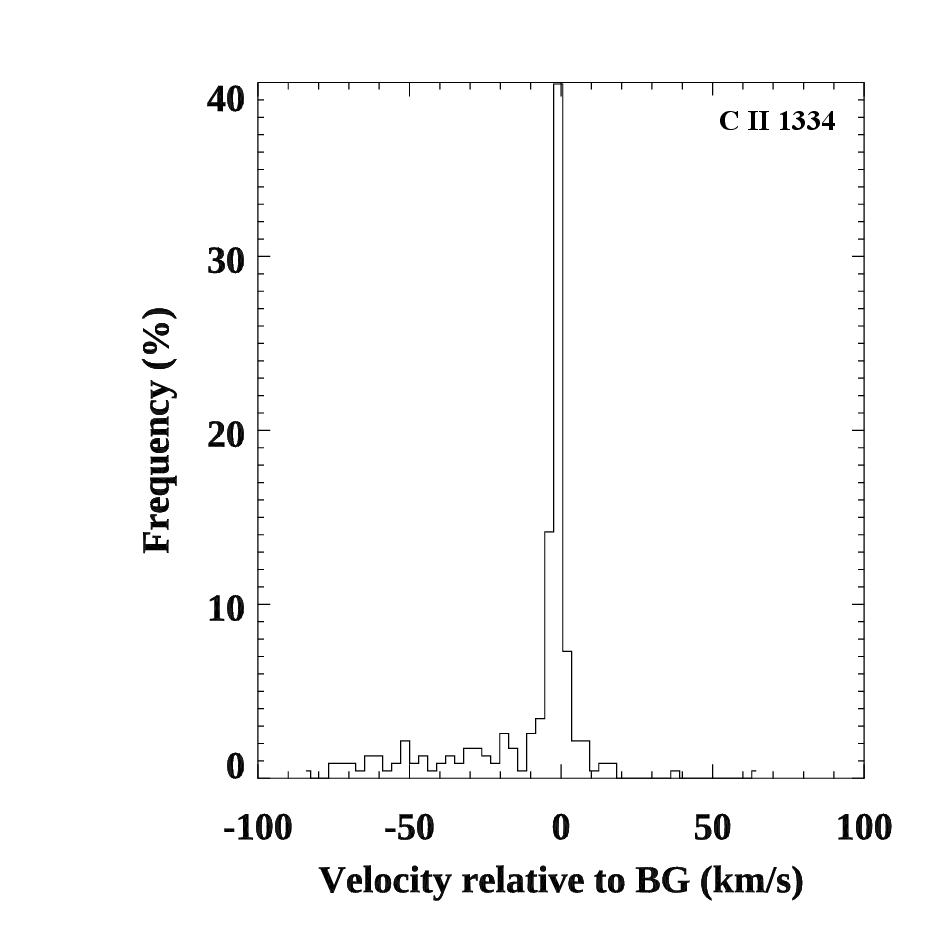}
    (b)\includegraphics[width=0.40\textwidth]{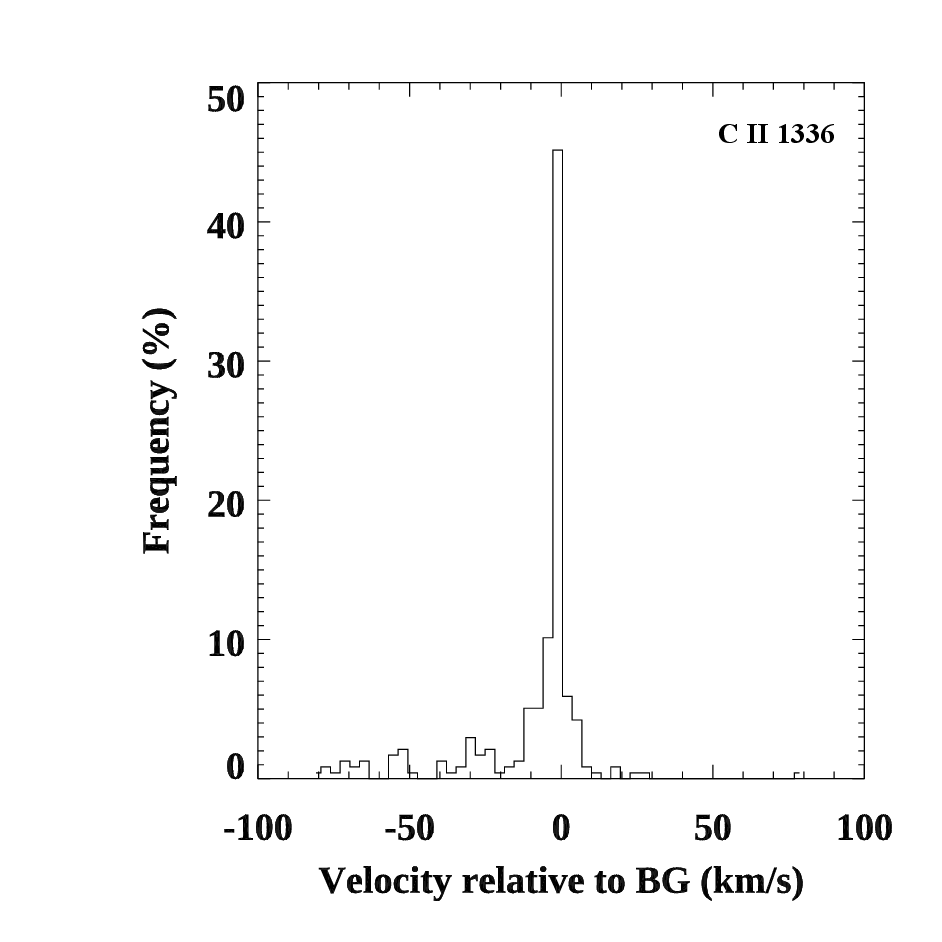}
    (c)\includegraphics[width=0.40\textwidth]{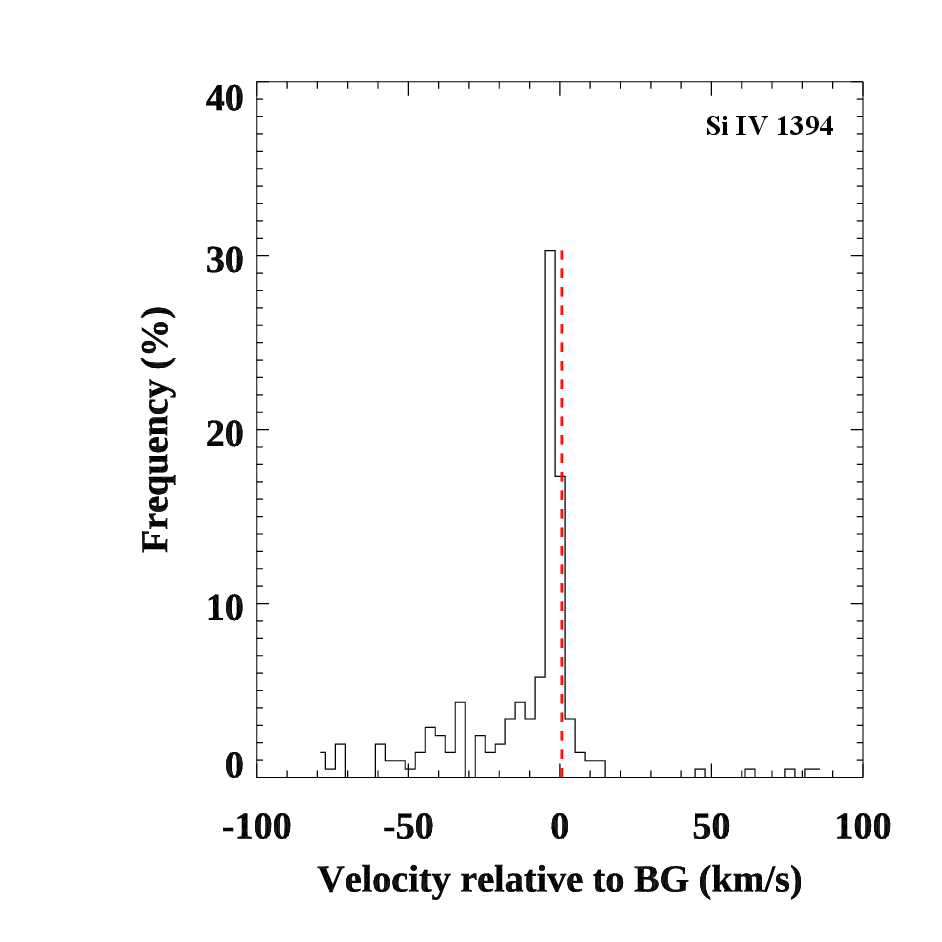}
    (d)\includegraphics[width=0.40\textwidth]{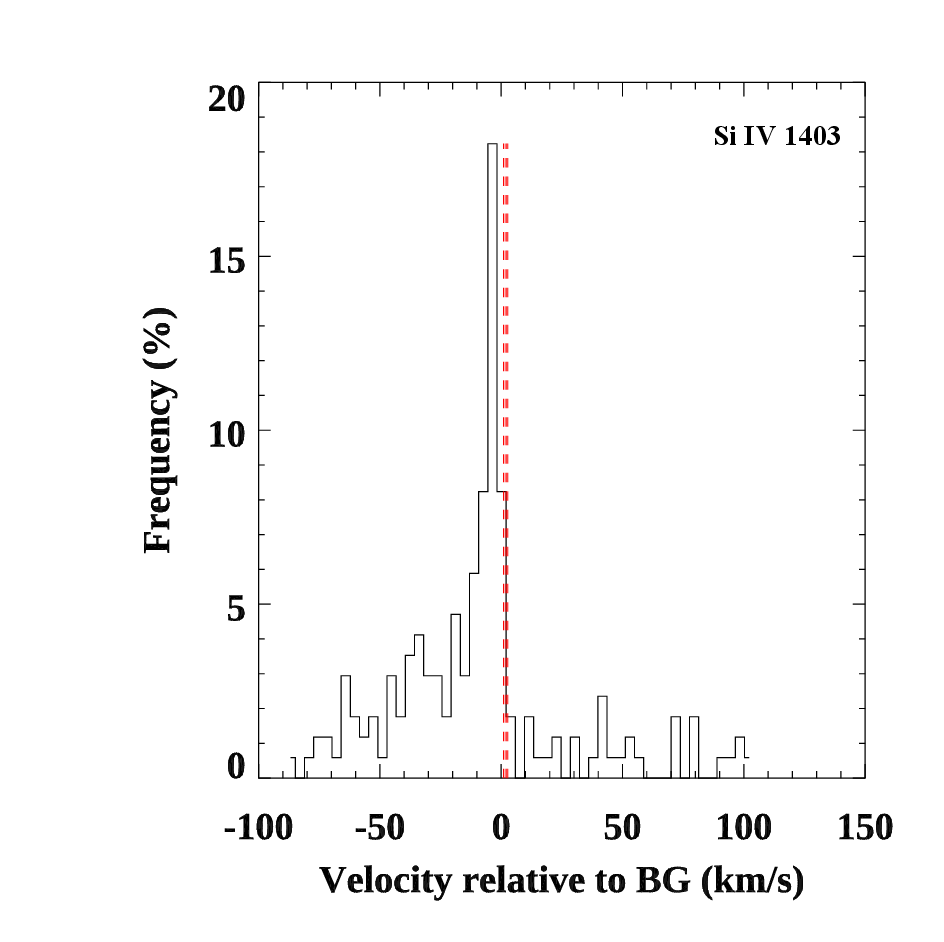}
    (e)\includegraphics[width=0.40\textwidth]{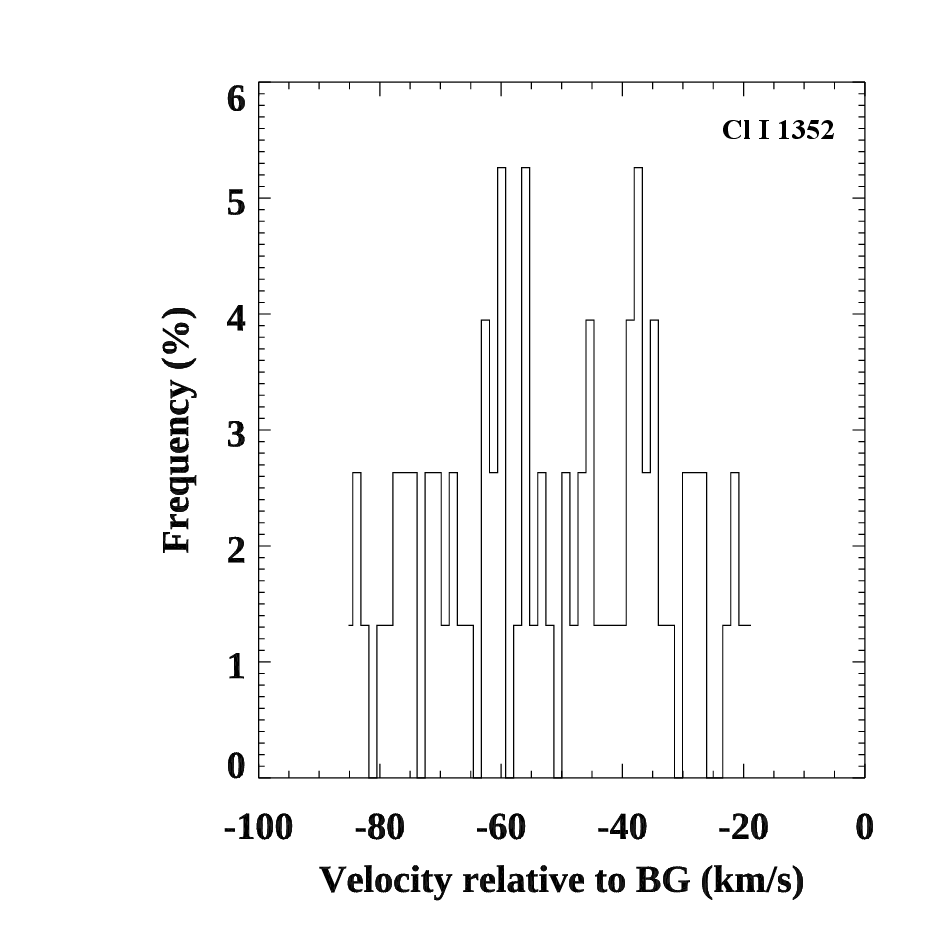}
    (f)\includegraphics[width=0.40\textwidth]{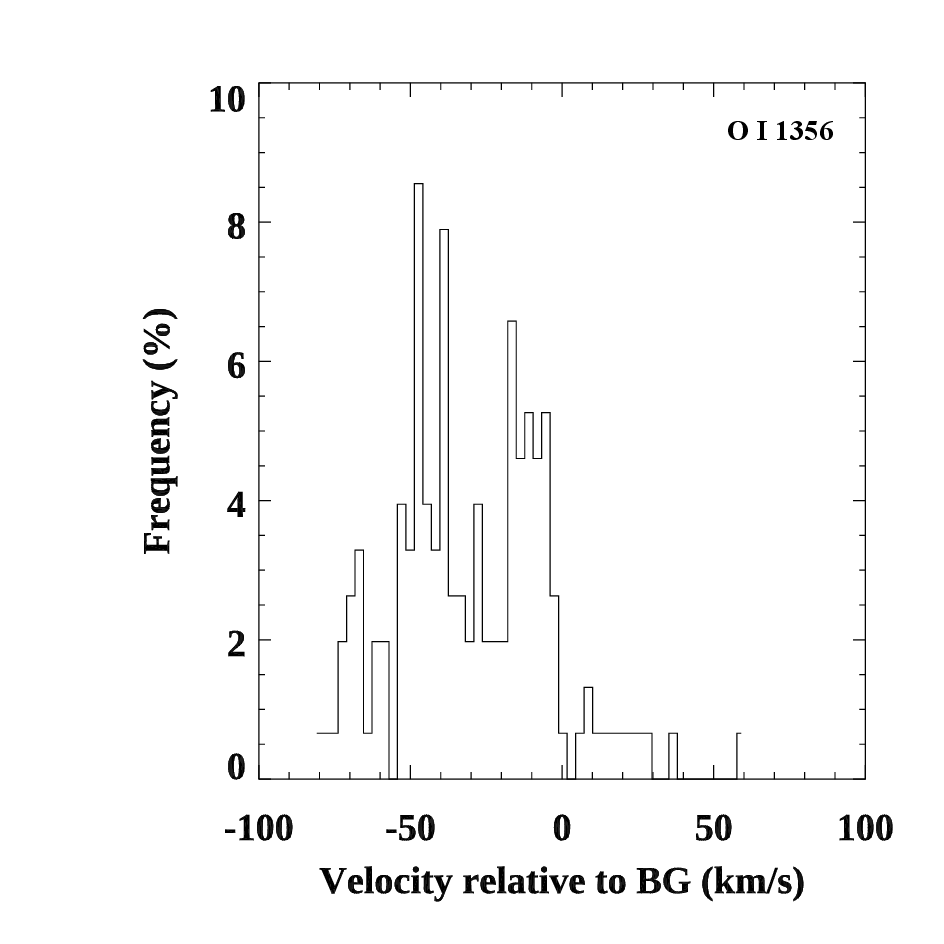}
    \caption{Histogram plots of measured spectroscopic velocities in (a) C {\sc{ii}} 1334, (b) C {\sc{ii}} 1336, (c) Si {\sc{iv}} 1394, (d) Si {\sc{iv}} 1403, (e) Cl {\sc{i}} 1352, and (f) O {\sc{i}} 1356, relative to the local background. BP velocity values are calculated using the mean of a small surrounding box, while the local background is calculated using the median value of a larger local box. Vertical dashed red lines indicate velocities of AQS BPs (if applicable).}
    \label{fig:LOS_vel_hist}
\end{figure*}

As this is a sit-and-stare observation, the IRIS slit remains in place rather than scanning across the FOV. A Si {\sc{iv}} 1403 \AA\ example of this ROI's spectroscopic data can be seen in figure \ref{fig:Si1403_vel_map}, whereby the green crosses indicate the positions at which BPs are detected as they cross the slit position - this number is severely reduced from the initial number of BP detections ($\sim12,600$) to $\sim150$. It is worth noting that this Doppler velocity map is cropped according to the time at which IRIS passes through the SAA (as previously mentioned) in the same manner that the imaging data is cropped. As mentioned previously, the reference central peak position for each wavelength is used for their corresponding calibrations as a precursory step to \textit{IRIS\_AUTO\_FIT}'s single (or double) Gaussian fittings.
LOS velocities of each successful BP spectroscopy detection are calculated by determining the mean of a box immediately surrounding the BP's detected position. The background velocity values are calculated by determining the median values within a slightly larger box surrounding the same detected BP position. The relative velocity of each BP is subsequently calculated by subtracting the initial velocity values from their corresponding background velocity values, which can be seen as histograms in figure \ref{fig:LOS_vel_hist} for (a) C {\sc{ii}} 1334, (b) C {\sc{ii}} 1336, (c) Si {\sc{iv}} 1394, (d) Si {\sc{iv}} 1403, (e) Cl {\sc{i}} 1352 and (f) O {\sc{i}} 1356.

Vertical dashed red lines indicate the relative velocity values for detected AQS BPs - most AQS BPs do not cross the slit and most of the spectroscopic data of those that do is unavailable. These histograms reveal that BPs demonstrate blue-shift, particularly in Si {\sc{iv}}, Cl {\sc{i}} and O {\sc{i}}. The intensity ratio of Si {\sc{iv}} 1394/1403 can be used as a diagnostic for opacity \citep{Tripathi_2020, Peter_2014}, whereby a ratio with a significant deviation from 2 may indicate a degree of optical thickness. It was found that the median of all Si {\sc{iv}} 1394/1403 pixel ratios for this dataset is $\sim1.9$ while the ratio of total intensity is $\sim2.4$, which suggests that the Si {\sc{iv}} lines are likely optically thin and therefore provide accurate velocity values in this instance.

\begin{figure*}[t]
    \centering
    (a)\includegraphics[width=0.40\textwidth]{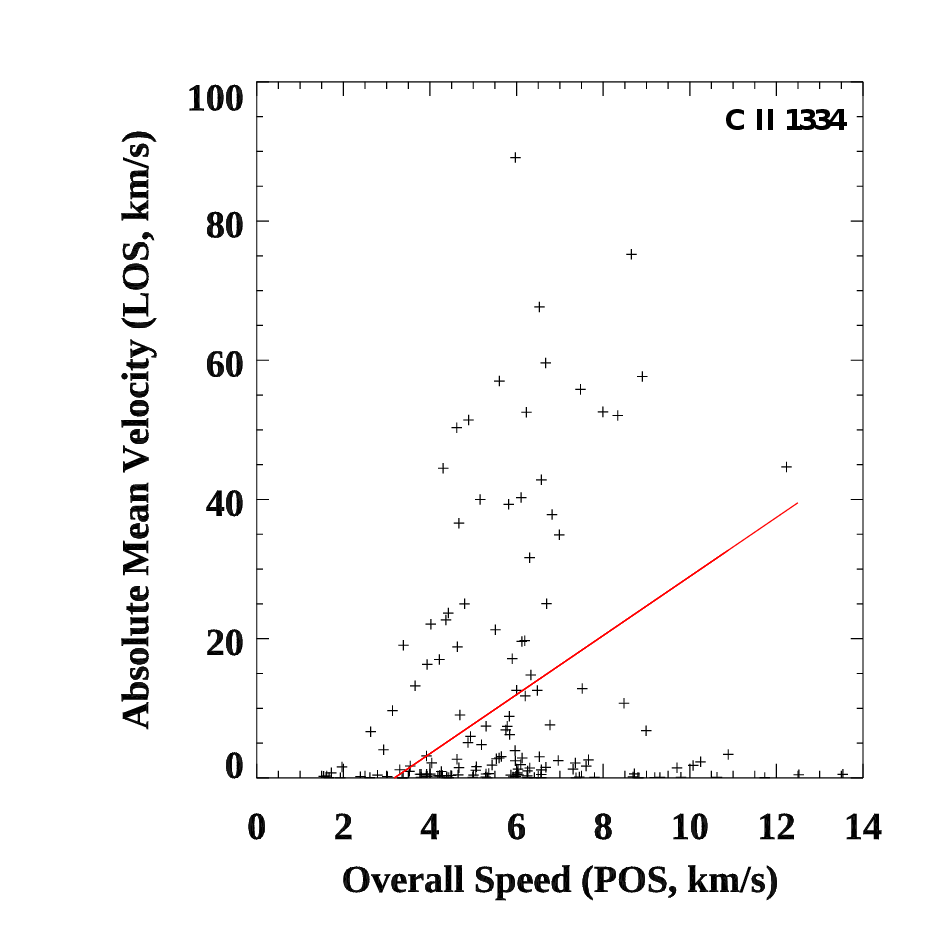}
    (b)\includegraphics[width=0.40\textwidth]{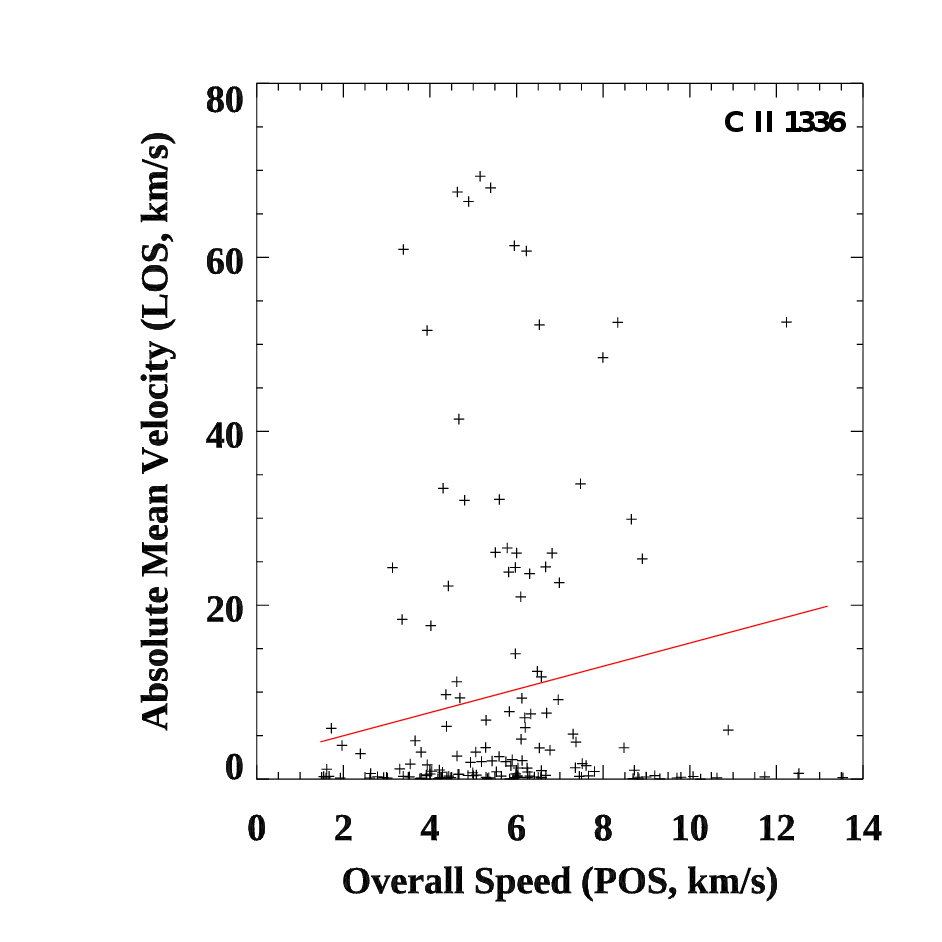}
    (c)\includegraphics[width=0.40\textwidth]{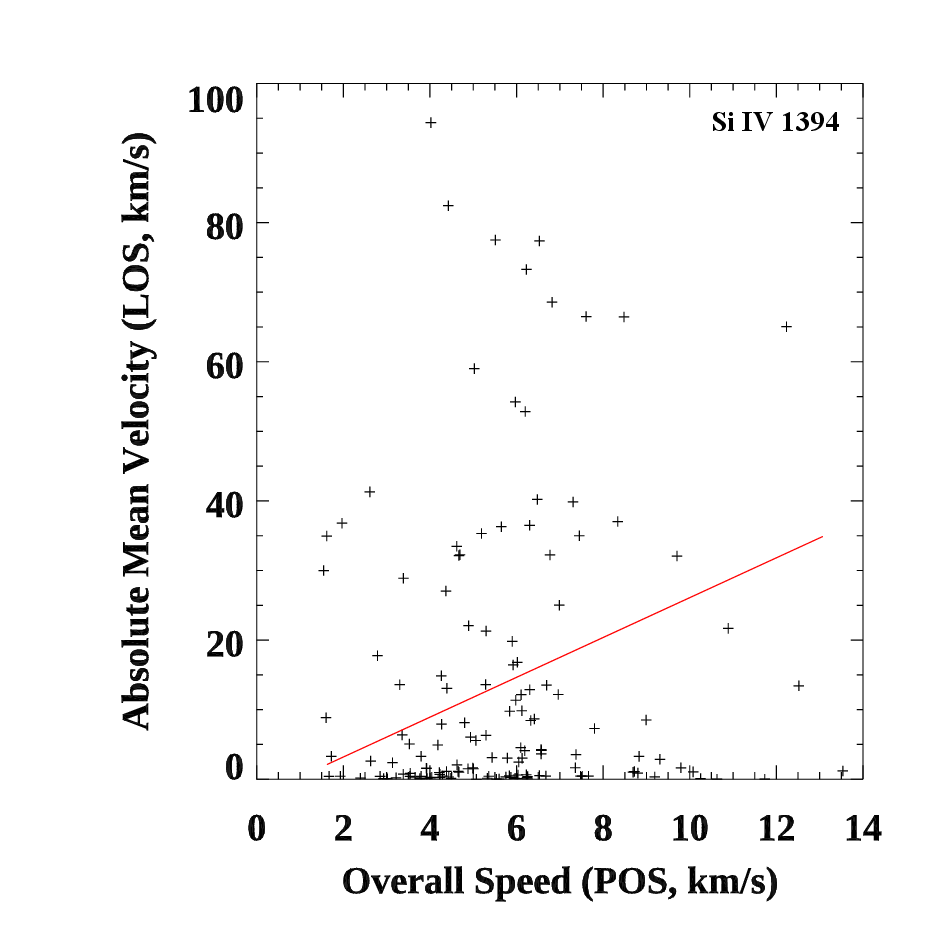}
    (d)\includegraphics[width=0.40\textwidth]{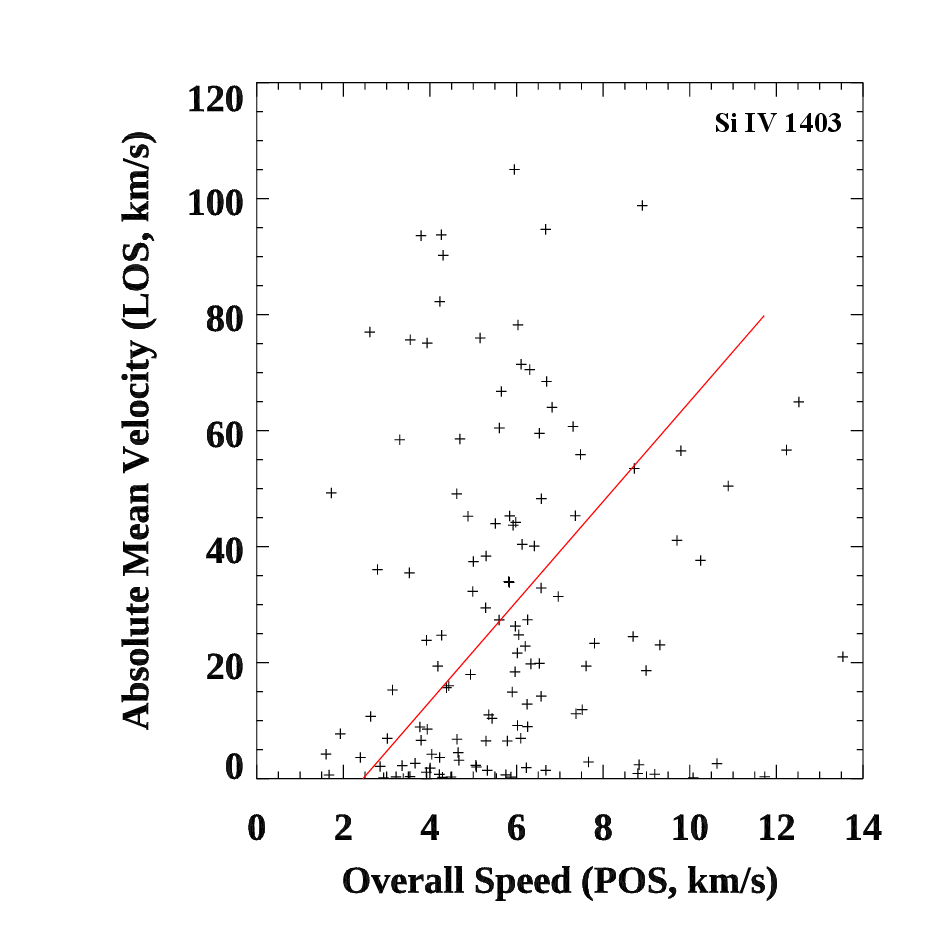}
    (e)\includegraphics[width=0.40\textwidth]{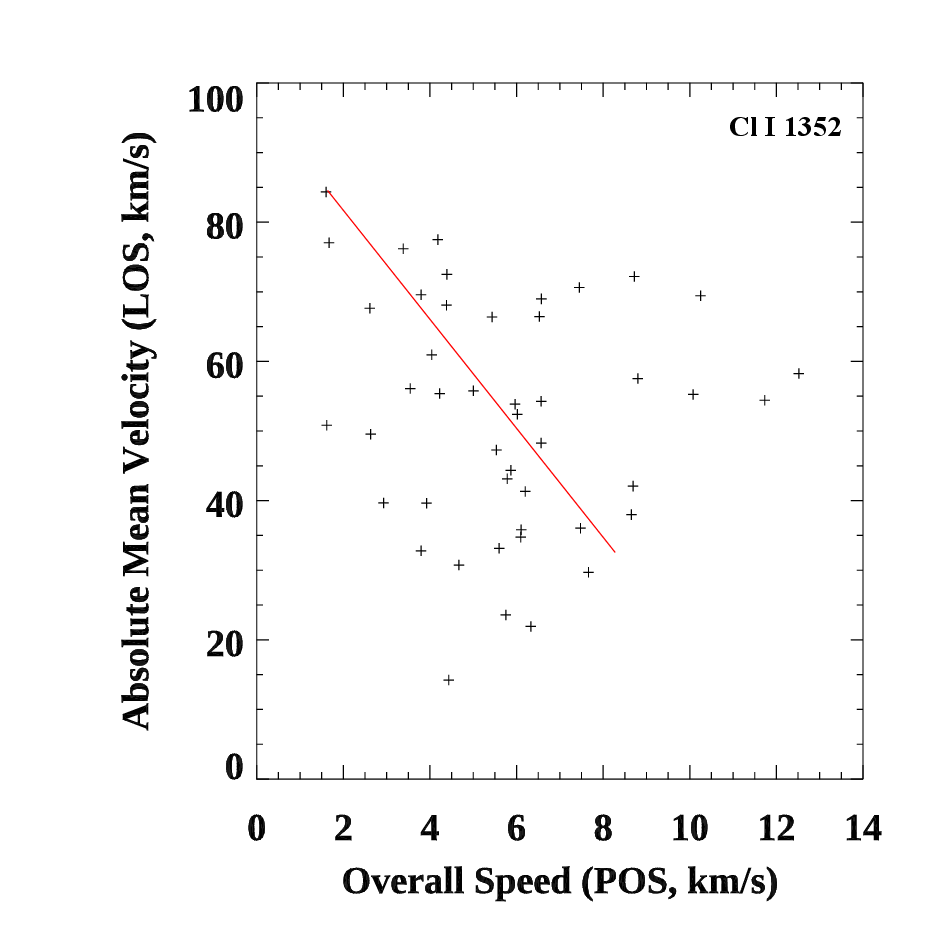}
    (f)\includegraphics[width=0.40\textwidth]{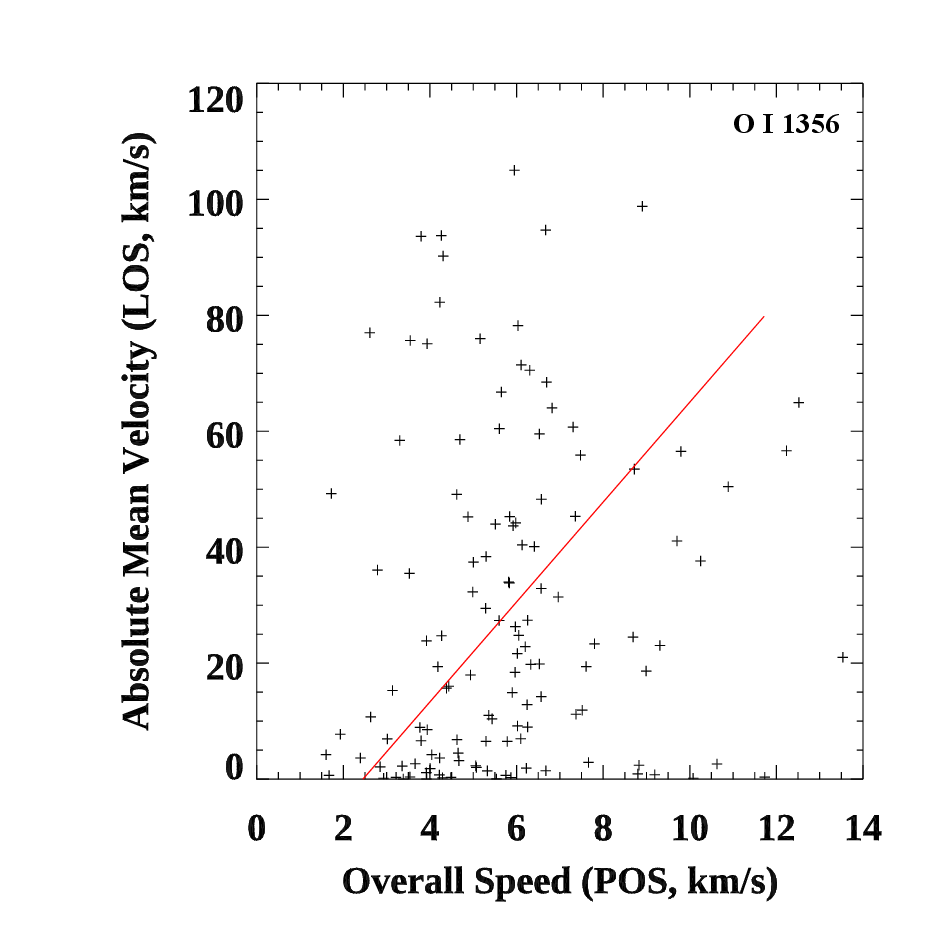}
    \caption{LOS Doppler BP velocities (absolute magnitude) vs corresponding POS overall speeds for (a) C {\sc{ii}} 1334, (b) C {\sc{ii}} 1336, (c) Si {\sc{iv}} 1394, (d) Si {\sc{iv}} 1403, (e) Cl {\sc{i}} 1352, and (f) O {\sc{ii}} 1356. Red lines indicate Deming regression fits.}
    \label{fig:LOS_vs_POS}
\end{figure*}

Figure \ref{fig:LOS_vel_hist}c-f also demonstrate some evidence of red-shift, suggesting a bidirectional nature. Figure \ref{fig:LOS_vel_hist}a-b largely demonstrates little to no BP LOS movement relative to their immediate background for over $40\%$ of BPs while the remaining BPs show some preference for blue-shift. C {\sc{ii}} 1336 is a doublet consisting of 1335.663 and 1335.708, which makes the curve non-symmetric \citep{Rathore_Carlsson_2015}. This may result in inaccurate velocity readings without an appropriate reference wavelength for each blended peak. However, this is beyond the scope of this current study and would require a subsequent deeper analysis of this particular line. Having said this, the velocity distributions of C {\sc{ii}} 1334 and C {\sc{ii}} 1336 are very similar (possibly due their similar formation heights), and \cite{Rathore_2015} demonstrate Pearson correlations of 0.69 and 0.63, respectively, at their formation height, suggesting that the C {\sc{ii}} 1334 velocities are accurate.

Cl {\sc{i}} 1352 becomes optically thick as the FOV moves away from disk centre. This dataset is centred at $X = 468\arcsec$, $Y = -194\arcsec$ which may have some effect on these relative velocity results. \cite{Lin_2015} show that O {\sc{i}} 1356 is typically optically thin and, from a Pearson correlation of 0.99, that O {\sc{i}} 1356 Doppler shifts provide a good diagnostic for LOS velocities.

Figure \ref{fig:LOS_vs_POS} demonstrates scatter plots of the (absolute) LOS Doppler velocities of each BP vs their corresponding detected overall POS speeds, whereby each panel matches the lines from figure \ref{fig:LOS_vel_hist}. Red lines indicate a Deming regression fit. There is a general trend of proportionality between Overall POS Speeds and Vertical LOS speeds, whereas (e) Cl {\sc{i}} is an exception with a distinct inverse relationship and a smaller number of LOS velocity results compared to the other wavelengths. Additionally, the LOS velocities are a magnitude higher than their corresponding POS speeds. Cl {\sc{i}}'s typically low opacity away from disk-centre may suggest erroneous LOS velocities (although this is currently unclear). It is worth noting that the POS speeds used in each panel are obtained entirely from Si {\sc{iv}} SJI data, therefore some discrepancies between LOS and POS are to be expected. We intend to include multi-wavelength POS statistics in a subsequent study in order to expand greatly on this type of analysis.

The gradient of the Deming fit for (a, b) both C {\sc{ii}} lines and (c) the Si {\sc{ii}} 1394 line are low as the majority of LOS speeds are fairly similar to their corresponding POS speeds or close to zero, with a few high-velocity outliers. The same cannot be said for (d) Si {\sc{iv}} 1403 and (f) O {\sc{i}} 1356, whereby a great deal of LOS speeds are far higher than their POS speeds. It is worth noting once more that the POS speeds are from Si {\sc{iv}} SJI data, and this comparison with O {\sc{i}} Doppler velocities doesn't necessarily represent a large POS vs LOS differential. Si {\sc{iv}} can occasionally undergo blending with the Ni {\sc{ii}} 1393.3 and CO 1393.5 lines which may explain the observed discrepancies in speed, although this has not been explored.

\section{Discussion}\label{sec:disc}

\subsection{POS speeds and interpretation}\label{sec:disc_interp}
An aspect of the magnetic canopy model of the solar atmosphere \citep{Jones_1985} is that the photosphere is arranged in a series of granules and supergranules (or network regions) interlaced with intergranular lanes. The model also suggests that magnetic field lines within the chromosphere are likely vertically orientated with respect to the surface everywhere (hereafter referred to as ``vertically oreintated" or VO) above these intergranular lanes, while the opposite is the case above network regions. 
A comparison of the IRIS dataset with the AIA image in figure \ref{fig:FOV_comp}(b) shows a correlation between the AQS domains across the IRIS FOV with the bright regions observed in 1700 \AA\ by AIA/SDO. This channel is sensitive to the temperature minimum of the photosphere and may also have contributions from TR emissions. This channel therefore shows enhanced intensity in the higher concentrations of magnetic field at network boundaries. 

Combining this IRIS/AIA comparison, the brightness results from table \ref{tab:data_AQS_vs_TQS}, figure \ref{fig:histograms_AQS_vs_TQS}, figure \ref{fig:2D-hist}, and the conclusion that AQS BPs are more energetic, we suggest that VO magnetic field lines within AQS domains (above intra-granular photopsheric lanes) restrict the POS travel distances of BPs detected therein compared to those detected in TQS domains (network regions) where magnetic field lines tend to be orientated more horizontally with respect to the surface (hereafter refereed to as ``horizontally orientated" or HO).

\begin{figure*}[t]
    \centering
    (a)\includegraphics[trim=0 -2cm 0 0,clip,width=0.45\textwidth]{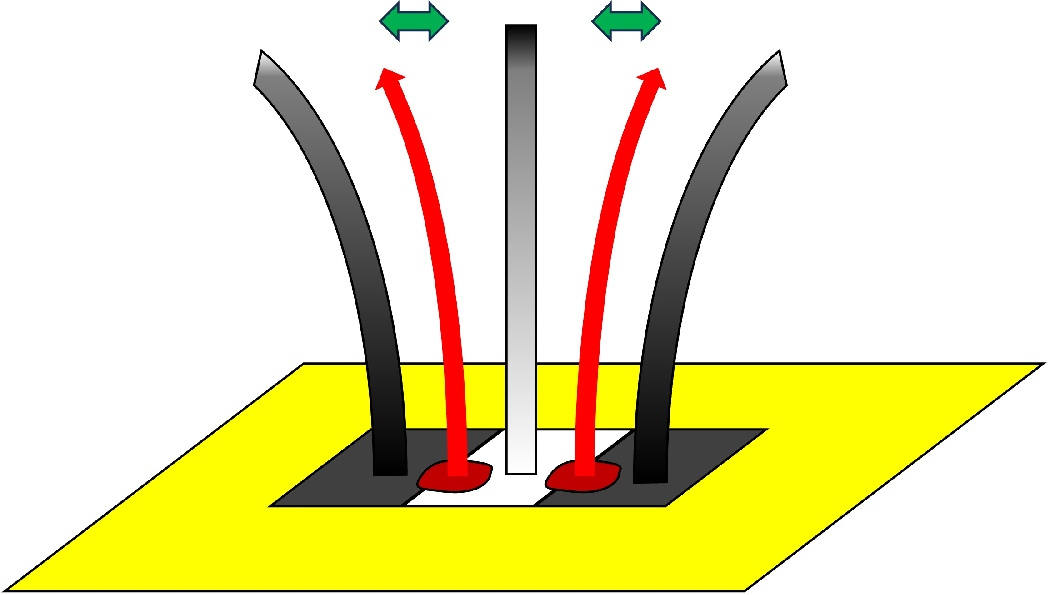}
    (b)\includegraphics[trim=0 -1.5cm 0 0,clip,width=0.45\textwidth]{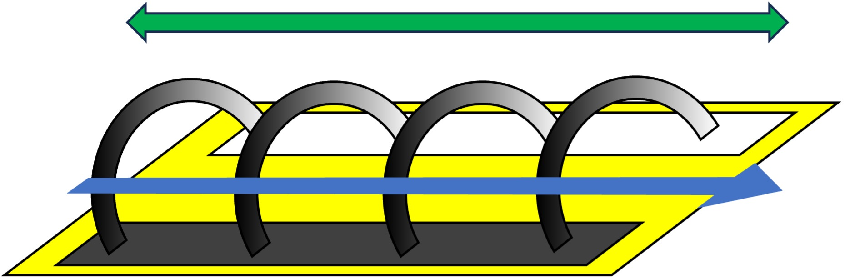}
    \caption{Schematic diagram illustrating the relationship between magnetic field structures and BP motions as they travel along current sheets within both the (a) AQS and (b) TQS domains. %(c) represents a top-down diagram of proposed BP paths within TQS domains. 
    Black and white regions represent opposite polarities. Red and blue arrows represent true motions of AQS and TQS brightenings, respectively. Green arrows represent magnitude of POS travel distances. See text for further details.}
    \label{fig:diagrams}
\end{figure*}

We speculatively interpret the results of this study as the following: these BPs are the result of small-scale magnetic reconnection events and can therefore follow current sheets instead of the magnetic field lines themselves; heated plasma as a result of reconnection can flow along current sheets if plasma pressure is sufficient between magnetic field lines of differing polarity. This FOV's chromospheric magnetic field structure influences the motion of these BPs according to the magnetic canopy model, and figure \ref{fig:diagrams} helps to demonstrate this. Dark/light regions represent magnetic fields of opposite polarity, and the shaded curves represent magnetic field gradients. VO AQS field lines restrict POS plasma flows, which is shown in \ref{fig:diagrams}a. The red arrows indicate BP plasma motion following photospheric emergence, and the green arrow demonstrates the relatively small POS motion. 
Conversely, figure \ref{fig:diagrams}b demonstrates larger TQS POS distances along HO magnetic fields, represented by the large green arrow and blue arrow.

As a continuation of our speculative interpretation, we suggest that BPs that occur closer to the photosphere – where there is greater magnetic energy and where magnetic fields tend to be more VO – tend to be brighter because the magnetic field is stronger near the photosphere. However, VO magnetic field lines result in a tendency for reduced POS motions. Conversely, BPs that occur higher in the atmosphere – where there is less magnetic energy and where magnetic fields tend to be more HO – are dimmer due to weaker magnetic fields, and these HO magnetic field lines allow greater POS travel distances. This offers an explanation for the inverse correlation between figure \ref{fig:2D-hist}g and \ref{fig:2D-hist}j's intrinsic brightness and overall travel distance results, and is reinforced by the histograms in figure \ref{fig:histograms_AQS_vs_TQS}, whereby AQS BPs are intrinsically brighter and also tend to move smaller POS distances. 
A possible avenue for future study would be to analyse the photospheric magnetic field using HMI data; if, for example, BPs from different domains prefer to follow motions parallel or perpendicular to magnetic field gradients, then this may both provide an explanation for BP motions as well as bolster the suggestion that the magnetic field structure (chromospheric and/or photospheric) has a significant effect on POS BP motions.

Perhaps then, these BPs may be considered UV bursts or IRIS bombs, which appear on spatial scales of 500-1000 km \citep{Young_2018} and a few arcseconds, respectively \citep{Peter_2014}. \cite{Young_2018}'s study demonstrates intensity variability throughout their UV burst's lifetimes, which is similar to that of \paperi's brief investigation of a similar event. Researching the variability of these BPs - as we have briefly done in this study by discussing the subtle differences in their average area and area OES - may grant further insight into the nature of these events.

If these BPs originate from the photosphere then they may share some similarities with spicules, whereby photospheric footpoint brightenings are often followed by plasma motion. The mechanism of rebounding shock fronts can give rise to an oscillating wake observable along the POS or line-of-sight \citep{Stein_1972}. The lower limit of spicule upward velocity of $\sim20$ km $s^{-1}$ \citep{Sterling_1988} is similar to those of BP overall POS speeds, and their durations of 3-5 minutes \citep{Tian_2018} also lie within the duration range of these BPs. However, spicules are generally categorised as dynamic jets of plasma, which implies a greater aspect ratio; no attempt to measure the dimensions of these BPs has been made other than the average area, but this may become another facet of study in future work. \cite{Hollweg_1982} postulates that, if spicules reach from photosphere to higher layers, then they must follow magnetic field lines with progressively reduced gravitational influence, i.e., non-vertical paths of motion. This model compliments the proposed mostly-vertical BP paths of figure \ref{fig:diagrams}a, provided that the spicule's path of motion becomes less inclined before reaching the TR. Less vertically inclined spicule paths may explain some of the larger TQS POS travel distances, although it is unclear as to whether spicule paths could explain the BPs's curved and complex motions.

We can conclude from this study that the chromosphere - perhaps most evident in figure \ref{fig:motion_AQS_vs_TQS} with its semicircular and criss-crossing BP motions - consists of very small-scale and complex magnetic and thermal structure. We can also conclude, based on the statistical and histogram results, that these BPs are likely magnetic in nature based on the magnetic canopy model of the chromosphere. 

There is a potentially tentative relationship between the distribution of these BPs, their behaviour, and the magnetic field orientation of the chromosphere (based on the correlation between the images in figure \ref{fig:FOV_comp}a and \ref{fig:FOV_comp}b with \ref{fig:FOV_comp}c). Our results warrant further investigation into other IRIS QS datsets, as well as those of more active regions. If these BPs originate from or near the photosphere and potentially reach TR temperatures (given the thermal range of IRIS's 1400 \AA\ channel), and given our postulation that these BPs may emit in other wavelengths, then it is worth investigating whether these BPs have contemporaneous detections in far hotter wavelength channels, such as 193 or 171 \AA\ from AIA which are sensitive to coronal temperatures. Previous studies using numerical modeling, such as that of \cite{Testa_2014} show that TR emissions in IRIS 1400\AA\ with spatial scales of $\sim5-10$ arcseconds and temporal scales of $\sim10$s can be reproduced through modelling of impulsive heating events following magnetic reconnection. These heating events can result in the acceleration of non-thermal electrons to coronal temperatures, which are also signatures typical of microflares. However, X-ray observations have found microflares to typically be $\sim20-30\arcsec$ in length and perhaps $\sim20-30\arcsec$ in thickness \citep{Hannah_2008, Glesener_2017, Cooper_2020}, which are far larger than any BP detected in this study. Complimentary Hinode and AIA observations may reveal whether there are particle accelerations present and coronal temperatures reached which could subsequently determine whether the characteristics of these BPs are closer to those of the fabled ``nanoflares" than their micro facsimile. Additionally, high resolution data from Solar Orbiter \citep[including the SPICE imaging spectrograph][]{Muller_2020} could provide supportive evidence for precursory reconnection events of BP observations.

It is possible that these BPs can be considered types of campfire events. Campfires exist 1000-5000 km above the photosphere, have a lifetime range of$\sim$1 minute to over 1 hour, and an average lengths/widths of $0.1-5.4$ Mm and $1.6 \pm 0.64$ Mm, respectively \citep{Berghmans_2021, Zhukov_2021}, which lie within the duration and area OES ranges (assuming an average circular BP shape with an area of $\sim0.17-2.88$ arcsec$^{2}$, equating to a diameter of $\sim0.35-1.41$ Mm). However, campfires have been known to show cool-plasma (dark) structures in EUV images \citep{Panesar_2021}, which does not appear to be the case for the FUV 1400 \AA\ results - the addition of AIA contemporaneous datasets and detections would help bolster the potential that these BPs are in fact campfires or not.

\subsection{Spectroscopy}\label{sec:disc_spec}

If our suggestion that AQS BP POS motions are restricted by VO magnetic field lines, then it is possible that AQS BPs would exhibit greater LOS speeds than TQS BPs. However, due to a lack of usable AQS BP spectroscopic data, an AQS vs TQS LOS velocity comparison remains inconclusive and will require the analysis another sit-and-stare dataset. The spectroscopic data does reveal TQS BPs to exhibit blue-shift and possibly some bi-directionality, implying that they do in fact move along the LOS.

The C {\sc{ii}} lines are typical of the upper chromosphere or low TR while the Si {\sc{iv}} lines are typical for the TR (at approximately 65,000K), and combined they provide a FUV diagnostic range of 3.7-5.2 log T \citep{Pontieu_2014}. Cl {\sc{i}} is formed in the mid chromosphere as a result of a fluorescence effect driven by the C {\sc{ii}} 1335.7A line \citep{Shine_1983}, while O {\sc{i}} 1356 forms throughout the whole chromosphere. Combined with the conclusions of \cite{Humphries_2021_b}, the multi-thermal nature of the BPs is evident. These BPs exhibit signatures in several slit-jaw imaging channels as well as several spectroscopic wavelengths. Photospheric spectroscopic data is not available for this dataset and, while we do not know at which atmospheric height these BPs originate from, the available spectra do reveal that the BPs likely lie within the chromosphere and either reach TR temperatures or even extend up into the TR. Additionally, it may be the case that these BPs do in fact reach TR heights; discrepancies between the LOS velocities of Si {\sc{iv}} and C {\sc{ii}} in figure \ref{fig:LOS_vs_POS} suggest that the BPs have accelerated as they progress from the high chromosphere into the TR. Speculatively, this may also be explained by the magnetic canopy model, as the suggestion is that magnetic field lines, while potentially more HO at lower altitudes, tend to align more vertically (with respect to the solar surface) at greater altitudes, potentially allowing for LOS acceleration. Or, potentially, as these are absolute LOS velocity values, BPs may begin their life in the low TR, fall at high speeds to the chromosphere and then slow down. Ideally, the progression of this type of combined spectroscopic and imaging analysis should focus on a QS sit-and-stare dataset at disk-centre, devoid of SAA contamination, that can facilitate as many wavelength channels as possible for multi-wavelength POS speed analyses for direct comparison with Doppler velocities from similar wavelengths.

\section{Conclusions} \label{sec:con}

We present the results of our general filter/threshold method for detecting small-scale brightenings in IRIS QS data. Applied to the 1400 \AA\ channel, we detect over 12,000 BPs, 11\% of which reside within more ``Active" domains, while 77\% reside within the surrounding ``Quiet" domain. The statistics and histogram plots of their characteristics are analysed and compared, revealing that, while area and durations differ very little between AQS and TQS detections, the AQS detections are intrinsically brighter, travel at slower POS speeds, and traverse shorter POS distances than their TQS counterparts. Polar plots of these BP motions reveal a North-South preference, and histogram plots reveal that, while AQS and TQS directions of motions are largely similar, BPs tend to follow linear rather than curved paths.
%In the AQS domain, the magnetic gradient along the path of motion is greater than the gradient perpendicular to the path of motion for the vast majority of BPs. Conversely, for the TQS, the vast majority of BP paths have perpendicular magnetic gradients greater than parallel.
%In the AQS domain, the magnetic gradient along the path of motion and the gradient perpendicular to the path of motion are largely equal for most BPs. Conversely, for the TQS, the majority of BP paths have perpendicular magnetic gradients greater than parallel.

%Having examined the underlying photosphere, a clear difference is determined between BPs from both domains, whereby the magnetic gradient along the path of motion is greater than the gradient perpendicular to the path of motion for the vast majority of AQS BPs, and vice versa for TQS BPs.
We postulate that these BPs are the result of magnetic reconnection events, and that these BP motions are the result of plasma following current sheets between regions of differing magnetic field polarity, rather than along magnetic field lines themselves. This explanation, through the use of schematic diagrams, provides an explanation for the BPs differing behaviour while complying with the magnetic canopy model of the chromosphere, as well as the photospheric inter- and intra-granular structures. 

Spectroscopic results cannot confirm any LOS velocity differences between AQS and TQS BPs but do reveal blue-shift and some bi-directionality in TQS BPs. These Doppler results suggest that BPs may move upwards from the chromosphere into the TR. A comparison of LOS absolute velocities and POS overall speeds demonstrate that most LOS and POS speeds are of a similar magnitude (albeit with several exceptions) but this analysis is limited in so far as POS speeds have only been extracted from Si IV 1400 SJI data.

Our next steps will be to expand this type of study to a large number of IRIS datasets, for both QS and AR region examples. We also intend to apply the detection code to simultaneous AIA datasets to determine whether these BPs approach coronal temperatures, and if they are viable candidates for contributing to coronal heating. Additionally, we intend to incorporate spectroscopic analyses of as many optically thin wavelength channels as possible and compare them with all SJI wavelength channels, if available.\\

We thank the anonymous referee for comments that greatly improved this work.
L.H. and D.K. acknowledge the Science and Technology Facilities Council (STFC) grant ST/W000865/1 to Aberystwyth University. H.M. acknowledges the STFC grant ST/S000518/1 to Aberystwyth University. D.K. acknowledges the Georgian Shota Rustaveli National Science Foundation project FR-22-7506. Special thanks to Mr Philip John Humphries for grammatical assistance. The SDO and HMI data used in this work are courtesy of NASA/SDO and the AIA, EVE, and HMI science teams. IRIS is a NASA small explorer mission developed and operated by LMSAL with mission operations executed at the NASA Ames Research center and major contributions to downlink communications funded by ESA and the Norwegian Space Centre. The authors wish to acknowledge FBAPS, Aberystwyth University for the provision of computing facilities and support.

\vspace{5mm}

\facilities{Aberystwyth University, IRIS
            }

\software{IDL
          }
\bibliography{main}
\bibliographystyle{aasjournal}

\end{document}